%% file: main.tex
\documentclass[10pt,twocolumn,twoside]{IEEEtran}
\IEEEoverridecommandlockouts 
\usepackage{url}
\usepackage{mathrsfs}
\usepackage{bbold}
\usepackage{graphicx,amstext,amsmath,amssymb,color,psfrag}
\usepackage{arydshln}
\usepackage{caption}
\usepackage{subcaption}

\usepackage{threeparttable}
\usepackage{booktabs}
\usepackage[hidelinks]{hyperref} 

\usepackage[latin1]{inputenc}
\usepackage{marvosym}
\renewcommand{\EUR}{\unskip~\text{\mvchr{164}}}
\DeclareInputText{128}{\EUR}
\usepackage{multicol}
\usepackage{booktabs, multirow} 
\usepackage{soul}
\usepackage[table]{xcolor} 
\usepackage{changepage,threeparttable} 
\usepackage{algorithm}
\usepackage{algpseudocode}
\usepackage{paralist}
\usepackage{environ}         
\usepackage{etoolbox}        
\usepackage{graphicx}        

\usepackage[]{times} 
\usepackage{float}
\usepackage{amsfonts}
\usepackage{placeins}
\usepackage{setspace}

\newcommand*\rot{\rotatebox{90}}

\include{commands}
\title{\BLUE{A game theory analysis of decentralized epidemic management with opinion dynamics }$^*$\thanks{$^*$This work  supported by ANR via the grant NICETWEET, number ANR-20-CE48-0009 and by CNRS via the grant COVEXIT.}}

\author{Olivier Lindamulage De Silva$^1$, Samson Lasaulce$^{2,1}$, Irinel-Constantin Mor\u{a}rescu$^{1,3}$, Vineeth S. Varma$^{1,3}$\thanks{$^1$ Universit\'e de Lorraine, CNRS, CRAN, F-54000 Nancy, France,\\ {\small \tt constantin.morarescu@univ-lorraine.fr}}\thanks{$^2$ Khalifa University, Abu Dhabi, UAE} \thanks{$^3$ Automation Department, Technical University of Cluj-Napoca}}

\begin{document}
\maketitle

\begin{abstract}
In this paper, we introduce a \BLUE{static} game that allows one to \BLUE{numerically} assess the loss of efficiency induced by decentralized control or management of a global epidemic. Each player represents a region which is assumed to choose its control to implement a tradeoff between socio-economic aspects and health aspects; \BLUE{the control comprises both epidemic control physical measures and influence actions on the region opinion.} The Generalized Nash equilibrium $(\mathrm{GNE})$ analysis of the proposed game model is conducted. \BLUE{The direct analysis of this game of practical interest is non-trivial but it turns out that one can construct an auxiliary game which allows one: to prove existence and uniqueness; to compute the GNE and the optimal centralized solution (sum-cost) of the game}. These results allow us to assess \BLUE{numerically} the loss (measured in terms of Price of Anarchy ($\mathrm{PoA}$)) induced by decentralization with or without taking into account the opinion dynamics.
\end{abstract}
\section{Introduction}\label{sec:intro}
In response to the outbreak of the Covid-19 epidemic in 2020, many countries adopted uniform centralized social distancing policies, such as China, France, Italy, and Spain, in an effort to contain the spread of the virus. However, this approach resulted in inadequacies between the severity level of the measures and the local situation, leading to negative consequences such as avoidable local economic losses, psychological damage, lack of acceptance from citizens, and frustration \cite{CHAUDHRY2020100464,bambra2020covid,MARTINSFILHO20214}. Decentralizing decision-making in a federal system presents several potential advantages, such as closer proximity to citizens, access to more accurate information, consideration of local needs and circumstances, improved economic performance, and increased public sector efficiency at the local level \cite{elazar1987exploring}. Thus, many countries have allowed regions
to adapt decision-making processes to local conditions, resulting in different public health guidance being implemented across regions within a given country.

This context motivates us to address the problem of decentralized epidemic management, which involves several interconnected geographical regions, such as countries, provinces or states. Each region has only local control over the epidemics and its own individual objectives. A central question is whether decentralization results in a significant performance loss in terms of a global efficiency measure. This problem is also relevant in economics,  when a company aims to maximize the dissemination of goods or services while delegating dissemination policies to local entities \cite{moruarescu2020space} or more generally for viral marketing \cite{OlivierECC2023}. 

Another important factor in epidemic propagation is the behavior of the individuals within the regions. \BLUE{On the one hand,} in response to the Covid-19 outbreak, people have spontaneously reduced social contact, stayed home whenever possible, adopted stricter hygiene or social distancing measures, or worn masks, regardless of government policy but following a fad. \BLUE{On the other hand, measures taken by governments have not always been followed, exhibiting behavioral drifts largely catalyzed by physical and digital social networks \cite{lasaulce2020efficient,Ndagnall2020Frontiers}. The effectiveness of epidemic management measures thus depends in part on its social acceptance. Therefore, it is important to couple them with the related opinion dynamics.}

\BLUE{Motivated by the above,} we propose a mathematical model to evaluate the effects of decentralization on epidemic management \BLUE{while taking into account the presence of the opinion dynamics of the regions.} We consider a relatively simple mathematical model that captures the main features of interest, consisting of a (generalized) strategic form game built from a networked Susceptible-Infected-Recovered (SIR) compartmental model \cite{magal2016final,mei2017dynamics} coupled with an opinion dynamics model  \cite{SBaikeTCNS2022,SBaikeCDC2021,RSebastianCDC2017,RSebastianTNSE2020,RSebastian2019ACC,lin2021discrete}. \BLUE{Note that in \cite{SBaikeTCNS2022,RSebastianTNSE2020} the goals pursued are different since on the one hand they don't consider a game scenario and on the other hand, the results focus on stability of the equilibrium points of the considered dynamics. To translate the interaction between the epidemic propagation and opinion dissemination across the network in a quantitative manner, they propose a concept known as the "Opinion-Dependent Reproduction Number".}

The proposed game considers each player as a geographical area aiming to minimize an individual cost, which implements a given trade-off between socio-economic losses, global/local losses in terms of the reproduction number of the virus \cite{preciado2014optimal,FLiuCDC2021,Anzum2020,gollier2020cost}, awareness costs, and a behavioral drift. The cost for each region depends not only on its action but also on the actions of neighboring regions through the epidemic propagation graph and opinion dynamics graph. We note that the proposed game model is a static or one-shot game model, where a player chooses a given epidemic local control action fixed over a finite time horizon, and \BLUE{a fixed number of} awareness campaigns are applied by each region to influence the beliefs of individuals in the social networks. We restrict our attention to the planning control problem of a single phase of an epidemic. Furthermore, each region is assumed to have its own virus transmission rate, and the propagation among regions is characterized by the cross-transmission rates. The efficiency loss associated with decisions concerning the health aspect is modeled by drift rates, and the population of each node recovers with a fixed recovery rate, depending on the capacity and performance of the health system \cite{hota2020closed}.

\BLUE{Compared to existing works (e.g., \cite{hota2019game2,omic2009protecting,hayel2014complete,trajanovski2015decentralized,LCSSO,HuangTCNS}) our main contributions are the following. First, we propose a static game over a networked SIR model coupled with a time-varying opinion dynamics model. 
Second, the paper sets a generalized strategic form of a static game that allows a tradeoff between key socio-economic and health aspects. We provide a complete analysis of the generalized Nash equilibrium (GNE). Note that the GNE accounts the existence of coupled constraints in the epidemic game, which was not addressed before. Third, it provides a thorough numerical analysis of the efficiency of decentralized management of epidemics through a popular efficiency measure: the Price of Anarchy (PoA)
.} The paper presents a detailed description of the model in Sec. II, a complete analysis of the corresponding GNE in Sec. III, and a numerical analysis of the game for a COVID-19-type scenario (Sec. IV).

\section{Problem statement}\label{sec:Problem-statement}

We consider a set of $K \geq 2 $ interconnected regions (e.g., countries, 
provinces, 
or states) that are affected by an epidemic; the region index is denoted by $k \in \mathcal{K}:=\{1,\ldots,K\}$. \BLUE{The time evolution of the epidemic of each region is governed by a SIR-type dynamical model  described in Sec.~\ref{subsec:epidemic-model}. The epidemic can spread from one region to another due to social interactions captured by the coupling between the dynamics within each region. Additionally, we assume the epidemic management is affected by a behavioral drift described by a linear opinion dynamics.} 
The epidemic management is assumed to be decentralized, which means that each region chooses the way the epidemic is mitigated or controlled over its own geographical territory. To model the underlying decision process, we propose a static game model whose strategic form is provided in Sec.~\ref{subsec:game-formulation}.

\subsection{Dynamical System Model} \label{subsec:epidemic-model}

\BLUE{In the sequel we use the following standard notations:\\[-5mm]
\begin{table}[h]
\centering
\begin{tabular}{c|l}
\textbf{Symbol} & \textbf{Description}  \\ 
\hline
 $s_k$ ; $s$ & fraction of susceptibles in Region $k$  \\
 $i_k$ ; $i$ & fraction of infected in Region $k$  \\
 $r_k$ ; $r$ & fraction of recovered in Region $k$  \\
$\beta_{k \ell}^0$ ; $B^0$ & natural virus transmission rate from $k$ to $\ell$ \\
 $\gamma_k$ ; $\gamma$  & removal/recovery rate within Region $k$  \\
$\widehat{\beta}_{k \ell}$ ; $\widehat B$  & maximum amplitude on $\beta_{k \ell}^0$ induced by OD \\
$u_{k\ell}$ ; $u$ & control policy of Region $k$ over Region $\ell$ \\
$\nu_{k \ell}$ ; $\nu$ & control by region $k$ on opinions from region $\ell$
\end{tabular}
\caption{Notations, the symbol after the semicolon represents the vector or matrix collecting the symbols before.}
\label{tab:notations}
\end{table}}

Note that $\frac{1}{\gamma_k}$ is called the average recovery period and 
$u_{k\ell} \in \mathcal{U}_{k\ell}$, $\mathcal{U}_{k\ell}:= [u_{k\ell}^\mathrm{min}, u_{k\ell}^{\max}]\subseteq[0,\beta_{k\ell}^0]$ and assumed to be constant over a time interval $[0,T]$, $T>0$. During the Covid-19 epidemics in 2020 control measures were typically constant over a period of a couple of weeks and updated from period to period; for this example of epidemics, choosing $u_{k\ell} = u_{k\ell}^{\max} $ would correspond to very severe lockdown and social distancing measures. The set where the control action $u_k=(u_{k1},\ldots,u_{kK})$ lies in is denoted by $\mathcal{U}_k =\prod\limits_{\ell\in\mathcal{K}}\mathcal{U}_{k \ell}$. Over a given time interval, the quantities $s_k\in[0,1]$, $i_k\in[0,1]$, and $r_k\in[0,1]$ evolve in continuous time and $t$ will be used as the corresponding time variable. Within each interval, each region is also allowed to implement influence control campaigns at given discrete time instants denoted by $t_n\in[0,T],\ n\in\{0,\ldots,N\}$, $N>1$, $t_{n+1} >t_n$. The opinion of Region $k$ is \BLUE{the aggregated/averaged value of the opinions in the region and it is assumed to evolve in a discrete-time manner and the opinion at time $t_n$ is denoted by $\theta_k(n) \in [0,1]$. The scalar quantity $\theta_k$ thus represents an abstraction of the global behavior of a region in terms of adhering (or not) to the control policy of the region.} The natural influence in terms of the opinion of Region $\ell$ on Region $k$ at time $t_n$ is assumed to follow a linear model and is represented by a weight $p_{k\ell}(n) \in [0,1]$. \BLUE{This weight captures the social interaction strength from Region $\ell$ on Region $k$. We also consider that each Region is able to adjust the social influence weight exerted by other Regions. Let us denote the control action on the social influence of Region $\ell$ on the influence from Region $k$ at instant $t_n$} by $v_{k\ell}(n) \in \mathcal{V}_{k\ell}$, $\mathcal{V}_{k\ell}:=[{v}_{k\ell}^{\min},{v}_{k\ell}^{\max}] \subseteq [0,1]$. For example, choosing $v_{k\ell}(n) = {v}_{k\ell}^{\min}$ would mean that Region $k$ reduces as much as possible the influence of Region $\ell$; in practice, this can be done by posting a large number of messages to counterbalance the influence of the other region or by simply applying information withholding. \BLUE{It is worth noting the asymmetry of the influence graph related both to the asymmetry of the social influence ($p_{k\ell}(n)\neq p_{\ell k}(n)$) and the independence of the actions ($v_{k\ell}(n)\neq v_{\ell k}(n)$).} By denoting ${v}_k(n)=({v}_{k1}(n),v_{k2}(n),\ldots,v_{kK}(n))$, the set where the control action $v_k =(v_k(0),\dots,v_k(N))$ lies in is $\mathcal{V}_k^{N+1}$, where $\mathcal{V}_k = \prod\limits_{\ell\in\mathcal{K}}\mathcal{V}_{k \ell} $. At last, we use the notations $\mathcal{N}_k$ and $\widehat{\mathcal{N}}_k(n)$ to respectively refer to the sets of neighbors of Region $k$ for the epidemic propagation and the influence propagation. The set of neighbors in the influence graph is allowed to vary over time. Some additional assumptions on the epidemic propagation and influence propagation graph will be added throughout the paper. 
The hybrid dynamics for the epidemic in Region $k$ in presence of interconnections and opinion dynamics can be written
$\forall (k,\ell)$, $ \forall n$,  $\forall t\in[t_n,t_{n+1})$, $\forall (u_{k\ell}, {v}_{k\ell}(n))\in \mathcal{U}_{k\ell} \times \mathcal{V}_{k\ell}$,
\begin{equation}\label{eq-sys-tot}
\hspace{-1em}\left\{\begin{array}{l}
\displaystyle{\frac{\mathrm{d} s_k}{\mathrm{dt} }}=-s_k(t)\sum_{\ell\in\mathcal{N}_k}\Big[ \beta_{k\ell}^0-u_{k\ell}+\theta_{k}(n) \widehat{\beta}_{k\ell} \Big]i_\ell(t),\\
\displaystyle{\frac{\mathrm{d} i_k}{\mathrm{dt} }}=-\frac{\mathrm{d} s_k}{\mathrm{d} t}-\gamma_k i_k(t),\\
\displaystyle{\frac{\mathrm{d} r_k}{\mathrm{dt}}}=\gamma_k i_k(t),\\
\displaystyle\theta_k(n+1)=\sum_{\ell\in \widehat{\mathcal{N}}_k(n)} {v}_{k\ell}(n){p}_{k\ell}(n)\theta_{\ell}(n)
\end{array}\right.
\end{equation}

\BLUE{\textit{Notation.} 
In addition to the commonly used notations given in Table 1, We also use the matrices: }
$\boldsymbol{D}_{\gamma} =\mathrm{Diag}(\gamma)$; $\boldsymbol{P}(n) = \left[{p}_{k\ell}(n)\right]_{1\leq k,\ell\leq K}$; the epidemic control action matrix $\boldsymbol{U}$ is defined by the entries ${U}_{k\ell} =\left\{\begin{array}{l}
     u_{k\ell}\text{ if }\ell\in\mathcal{N}_k\\
     0\text{ otherwise}
\end{array}\right.$;\\ the influence control action matrix at time $t_n$ is defined by the entries ${V}_{k\ell}(n) =\left\{\begin{array}{l}
     v_{k\ell}(n)\text{ if }\ell\in \widehat{\mathcal{N}}_k(n)\\
     0\text{ otherwise}
\end{array}.\right.$ \BLUE{The symbol $\odot$ denotes the Hadamard (element-wise) product.}\hfill$\Box$\\ 
With these notations, the system dynamics rewrites in the following compact form:
$\forall t\in[t_n,t_{n+1}),\ n\in\{0,\ldots,N\}$,
\begin{align}\label{eq-Matrixdyn}
\hspace{-3mm}\left\{\begin{array}{l}
    \displaystyle\frac{\mathrm{d} s}{\mathrm{dt}}= - \mathrm{ \mathrm{Diag}}(s(t))\left[\boldsymbol{{B}^0}-\boldsymbol{U}+\mathrm{Diag}(\theta(n))\widehat{\boldsymbol{B}}\right] i(t)\\
    \displaystyle\frac{\mathrm{d} i}{\mathrm{dt} }= -  \displaystyle\frac{\mathrm{d} s}{\mathrm{dt}}- \boldsymbol{D}_{\gamma} i(t)\\
    \displaystyle\frac{\mathrm{d} r}{\mathrm{dt} }=\boldsymbol{D}_{\gamma} i(t),\\
    \theta(n+1)=\left[\boldsymbol{V}(n)\odot\boldsymbol{P}(n)\right]\theta(n).\\
\end{array}\right. 
\end{align}
To conclude the presentation of the considered dynamical model, several mild conditions are assumed to be met. \begin{ass}\label{assumption}
\noindent (i): $\forall k,\ell,\ \beta_{k\ell}^0=0\iff \widehat{\beta}_{k\ell}=0$.\\
\noindent (ii) $\forall n\in\{0,\ldots,N\}$, $\boldsymbol{P}(n)$ is a row-stochastic matrix.\\ 
\noindent (iii): $\forall n\in\{0,\ldots,N+1\}$, the matrix $\boldsymbol{D}_{\gamma}^{-1}[\boldsymbol{B}^0-\boldsymbol{U}+\mathrm{Diag}(\theta(n)) \widehat{\boldsymbol{B}}]$ is non-negative and irreducible.
\hfill$\Box$
\end{ass}
Condition (i) means that if the virus is not physically transmitted between two regions, it is also not transmitted through a change in behavior between the two regions and vice-versa. Condition (ii) \BLUE{states that the uncontrolled opinion dynamics follows a very standard consensus model. While this choice is often made in the literature, it may not accurately represent some real dynamics over social networks.} 
Condition (iii) is verified when the controlled epidemic graph is strongly connected. This condition is reasonable since physical interactions are well-developed between many geographical regions.

\subsection{Generalized Strategic Form Game Model}\label{subsec:game-formulation}

 The first equation of (\ref{eq-sys-tot}) shows that the fraction of susceptibles in Region $k$ depends on the fraction of infected in the neighboring regions. Therefore the control actions of the neighbors of Region $k$ impact what happens in Region $k$ and thus its decision yielding a game. 
The most simple mathematical model for a game is given by the strategic form game model (see e.g., \cite{lasaulce2011game}) which 
comprises three components: the set of players, the sets of strategies, and the players' cost functions. When 
each player has a range of actions that depends on the actions of other players one needs to add one more component, the set of coupled constraints. This model with four components is called the generalized strategic form (see e.g., \cite{arrow1954existence,dutang2013existence}). Let us first describe the three conventional components and then 
we introduce the set of coupled constraints. The set of players here is the set of regions $\mathcal{K} = \{1,\ldots,K\}$ and
the sets of strategies coincide with the set of actions. The action of Region $k$ is given by the vector $(u_k,v_k)$ that is, the set of its actions is $\mathcal{U}_k \times \mathcal{V}_k$. The cost function of a player is chosen to be a tradeoff between a cost associated with the control actions, the local virus reproduction number, the global virus reproduction number, and a loss term due to the perturbation induced by the opinion. First, we provide the expression of the cost function for each Region $k$ and then we give some explanations about its construction: 
\begin{equation}
\begin{array}{ll}\label{eq:costgame1}
     &  \hspace{-2em} J_k(u,v):=-a_k  \displaystyle{\sum_{\ell\in\mathcal{N}_k}} \log\left(1 -\frac{u_{k\ell}}{\beta_{k\ell}^0}\right)\\
     &  \hspace{-2em} + b_k^{\mathrm{local}} \displaystyle{ {\sum_{n=0}^{N+1} \sum_{\ell\in\mathcal{N}_k}} \frac{\beta_{k\ell}^0-u_{k\ell}+ \theta_k(n) \widehat{\beta}_{k\ell}}{\gamma_k}}\\
 &\hspace{-2em}+ b_k^{\mathrm{global}}\displaystyle{{\sum_{n=0}^{N+1}}\rho\left(\boldsymbol{{D}}_{\gamma}^{-1} \left(\boldsymbol{{B}^0}-\boldsymbol{U}+\mathrm{Diag}(\theta(n))\widehat{\boldsymbol{B}}\right)\right)}\\
\end{array}
\end{equation}
\begin{equation*}
\begin{array}{ll}
 &\hspace{-2em} - {c_k \displaystyle{\sum_{n=0}^N\sum_{\ell\in \widehat{\mathcal{N}}_k(n)}}  \log({v}_{k\ell}(n))} + {d_k \displaystyle{ \sum_{n=0}^{N+1}} \theta_k(n)},
\end{array}    
\end{equation*}
\noindent 
where $(a_k, b_k^{\mathrm{local}},b_k^{\mathrm{global}}, c_k,d_k) \in \Rlo^5$ are non-negative parameters and $\rho(\boldsymbol{M})$ stands for the spectral radius (i.e., the largest modulus of an eigenvalue) of the matrix $\boldsymbol{M}$. Note that the reproduction rate terms are important and are often considered in the literature of epidemics (see for instance [18-20]) but only in the presence of a single decision-maker setting, thus not for a game. The other terms are new and are better motivated next.

\textit{Remark 1.} 
A common choice for the cost associated with the control action (namely, the first and fourth terms of $J_k$) is to assume a monotonic linear or quadratic expression (see e.g., \cite[Section 2.2.2]{charpentier2020covid}\cite{LCSSO}). Here we assume a logarithmic cost which not only allows one to still have a smooth, monotonic, and convex cost but also offers some posynomiality property that facilitates the non-trivial analysis of the GNE of the game. Interestingly, for some typical ranges for the control actions as those used for the Covid-19 case, the approximation of the log function by a linear function is very reasonable. For instance, when $u_{k\ell}\leq 0.53\beta_{k\ell}^0$ (or $v_{k\ell}(n)\geq 0.53$) the relative difference between $-\log(\frac{\beta_{k\ell}^0-u_{k\ell}}{\beta_{k\ell}^0})$ and $\frac{u_{k\ell}}{\beta_{k\ell}^0}$ (or $-\log(v_{k\ell}(n))$ and $-v_{k\ell}(n)$) is less than $30\%$. In other words, by restricting the action space of each player, one can assume that considering the logarithmic form to penalize the control action is equivalent to the linear one.

\textit{Remark 2.} The second and third terms of $J_k$ can respectively be interpreted as a local reproduction number (see \cite{gollier2020cost}) and a global reproduction number (see \cite{FLiuCDC2021}). We recall that (see \cite{FLiuCDC2021}), if the global reproduction number $\rho\left(\boldsymbol{{D}}_{\gamma}^{-1} \left(\boldsymbol{{B}^0}-\boldsymbol{U}+\mathrm{Diag}(\theta(n))\widehat{\boldsymbol{B}}\right)\right)$ is strictly less than $1$, the epidemic dies out in all the regions. Depending on the values of $b_k^{\mathrm{local}}$ and $b_k^{\mathrm{global}}$ a region will make the trade-off between 
the local and the global situation of the epidemics. \BLUE{The last term of the cost function accounts for the cost of the mismatch between public opinion and the policy of the region, which is not desirable for the latter. In the case where people follow the rules, the corresponding cost is small whereas it increases (linearly for simplicity) as people do not comply. This term might be neglected in practice e.g., when the associated (say monetary or health) cost can be neglected.} Additionally, motivated by practical considerations such as those encountered with the management of Covid-19 epidemics, we assume the existence of a set of constraints which includes a coupled constraint (in the sense of Rosen \cite{rosen1965existence}) on the game. The game action profile $(u,v)$ has to meet the following constraints: $(u,v)\in\mathcal{C}:=\prod\limits_{k=1}^K\mathcal{C}_k(u_{-k},v_{-k})$ where \color{black}$\mathcal{C}_k(u_{-k},v_{-k}):= \Big\{(u_k,v_k)\in\mathcal{U}_k\times\mathcal{V}_k:\ \forall n\in\{0,\ldots,N\},\ m\in\{0,\ldots,N+1\}$ \begin{eqnarray}
    &&\displaystyle{\sum_{\ell\in\mathcal{N}_k}} 
   \frac{u_{k\ell}}{\beta_{k\ell}^0} \leq \phi_{k},\ \displaystyle{\sum_{\ell\in \widehat{\mathcal{N}}_k(n)}} \frac{1}{{v}_{k\ell}(n)} \geq \widehat{\phi}_{k}(n),\label{eq:Constraint_general}\\
    && \displaystyle{\sum_{\ell\in\mathcal{N}_k}}\frac{\beta_{k\ell}^0-u_{k\ell}+ \widehat{\beta}_{k\ell} \theta_{k}({m})}{\gamma_k} \leq \mathrm{R}_k^{\max},\ \theta_k(m)\leq \theta_k^{\max}\Big\}.\notag
\end{eqnarray}
 At this point, some comments on the construction of the cost functions and the additional set of constraints are in order.\\ 

\textit{Remark 3.} We have added two budget constraints on the control actions $u_k$ and $v_k$. Notice that these individual constraints could have been directly integrated into the definition of the action sets for the players. However, the structure of the budget constraint on $v_k$ is easier to be understood after knowing about the cost function structure. Indeed, the constraint $\displaystyle{\sum_{\ell\in \widehat{\mathcal{N}}_k(n)}} \frac{1}{{v}_{k\ell}(n)} \geq \widehat{\phi}_{k}(n)$ can be rewritten, with a change of variable, as $\displaystyle{-\sum_{\ell\in \widehat{\mathcal{N}}_k(n)}} \log({v}_{k\ell}(n)) \leq \widehat{\psi}_{k}(n)$. At last, note that the constraints on the local reproduction numbers and those on $\theta_k(m)$ are coupled constraints because of the presence of $\theta_k(m)$, which leads us to consider the GNE as a suitable solution concept for the considered game. Finally, the generalized strategic form of the game when integrating all the constraints writes as:
\begin{equation}\label{eq-game}
   { \mathcal{G}:=\Big(\mathcal{K}, \Big(\mathcal{U}_k \times \mathcal{V}_k \Big)_{ 1\leq k\leq K},   \Big(\mathcal{C}_k \Big)_{1\leq k\leq K},\big({J}_k\big)_{1\leq k\leq K}\Big)}.
\end{equation} 

 \section{Generalized Nash Equilibrium Analysis}
 \label{sec:Main}

\BLUE{Because of the presence of a coupled constraint (motivated by practical considerations), the conventional NE cannot be retained a a solution concept. This is why we resort to a more involved solution concept namely, the GNE.} A GNE for the generalized strategic form game $\mathcal{G}$ is defined as follows. 

 \begin{defn} A GNE for the game $\mathcal{G}$ is a point $(u^\star, v^\star)$ such that $\forall k \in \mathcal{K}$,
 \begin{equation}
     (u_k^\star, v_k^\star) \in \argmin_{(u_k, v_k) \in \mathcal{C}_k(u_{-k}^\star, v_{-k}^\star)} J_k(u_k,v_k, u_{-k}^\star, v_{-k}^\star ).
 \end{equation}
 \end{defn}

A fundamental issue for the equilibrium analysis is the existence issue. There are useful existence theorems for strategic form games whose cost functions are individually convex or quasi-convex (see e.g., \cite{lasaulce2011game}). Such geometrical properties are not available here, which makes the existence analysis non-trivial and not a special case of existing general theorems. Remarkably, it turns out to be possible to construct an auxiliary game whose existence property guarantees, by equivalence, the existence of an equilibrium in $\mathcal{G}$. The auxiliary game even allows the uniqueness issue to be treated and to build an algorithm to determine the unique NE of $\mathcal{G}$. In addition to conducting the equilibrium analysis in this section (existence, uniqueness, determination), we also provide the equilibrium efficiency measures retained for the numerical analysis section. To facilitate the reading and make the results easy to exploit, the choice made by the authors is to state here only the derived results and to provide all the technical aspects and details in the Appendix section (Appendix-A).    

\subsection{Existence and uniqueness analysis}\label{subsec:existence_uniqueness}

To prove the existence and uniqueness of a GNE in $\mathcal{G}$, first, it is assumed that the less trivial term of $J_k$ is always present that is,  $\forall k,\ b_k^{\mathrm{global}}>0$. Second, \BLUE{since the game has no obvious geometrical properties such as convexity or quasi-convexity which would facilitate its analysis,} we introduce an auxiliary game $\widetilde{\mathcal{G}}$ which is obtained from $\mathcal{G}$ by performing appropriate changes of variables. The rationale for making these changes of variables is to \BLUE{exhibit a posynomiality property of} the opinion state $\theta_m(n)$ w.r.t. the influence control action $v_k$ (see Appendix-A). These changes of variables are as follows: $\forall n\in\{0,\ldots,N\}$ and $\forall (k,\ell_n,\ldots,\ell_0)\in\mathcal{K}^{n+2}$, $\omega_{k\ell_n\ldots\ell_0}(n)={v}_{k\ell_n}(n)\times{v}_{\ell_n\ell_{n-1}}(n-1)\times\ldots\times{v}_{\ell_{1}\ell_0}(0)$; $\xi_{\omega_{k\ell_n\ldots\ell_0}}(n)=\log{(\omega_{k\ell_n\ldots\ell_0}(n))}$; $\xi_{y_{k\ell}}=\log{(\beta_{k\ell}^0-u_{k\ell})}$. The auxiliary game $\widetilde{\mathcal{G}}$ has the following form: 
\begin{equation}
\widetilde{\mathcal{G}}=\Big(\mathcal{K}\cup\{K+1\}, \Big(\boldsymbol{\Pi}_{k} \widetilde{\mathcal{C}}\Big)_{1\leq k\leq K+1},\big(\widetilde{J}_k\big)_{1\leq k\leq K+1}\Big)
\end{equation}
where $\mathcal{K}\cup\{K+1\}$ represents the set of auxiliary players; $\widetilde{J}_k$ corresponds to the auxiliary individual cost functions given in \eqref{eq:tildeJk}; $\boldsymbol{\Pi}_{k} \widetilde{\mathcal{C}}$ is the projection of the coupled constraint set $\widetilde{C}$ on the action vector of the $k^{\text{th}}$-player in \eqref{eq:tildeC}. Exploiting the introduced auxiliary game, we have the following result.

\begin{prop}\label{Prop:Existence_Uniqueness}
 If $b_k^{\mathrm{global}}>0, \forall k$, the game $\mathcal{G}$ possesses a unique $\mathrm{GNE}$; which is denoted by $(u^\star, v^\star)$.\hfill$\Box$
 \end{prop}
 \noindent\emph{$\mathbf{Proof.}$} The proof is provided in Appendix-B. Therein, it is proved that a GNE in $\mathcal{G}$ becomes, by change of variables, a GNE of $\widetilde{\mathcal{G}}$ and conversely. One then proves that there exists a unique GNE in $\widetilde{\mathcal{G}}$.

\subsection{Efficiency measures}\label{subsec:EffectivenessofGNE}

One of the main objectives of this paper is to assess the potential inefficiencies that might be induced by decentralizing the management or control of an epidemic. A famous and well-used measure of global efficiency is given by the Price of Anarchy ($\mathrm{PoA}$) of a game \cite{papadimitriou2001algorithms}. For the sake of clarity, let us introduce the sum-cost function $J = \displaystyle{\sum_{k=1}^K} J_k$. To refine our efficiency analysis, we not only consider the original version of the PoA (which is denoted by $\mathrm{PoA}_{uv}$) but also two useful variants of it:
\begin{equation}\label{eq-POAuv}
\mathrm{PoA}_{uv} = \frac{J( u^\star, v^\star)}{\min\limits_{(u,{v})\in\mathcal{C}} J(u,{v})}
\end{equation} 
in which both $u$ and $v$ are controlled partially by the players and the uniqueness result is exploited;\\
\begin{equation}\label{eq-POAu}
\mathrm{PoA}_{u} = \frac{J(u^\star(1_{K^2(N+1)}), 1_{K^2(N+1)} )}{\min\limits_{(u,{v})\in\mathcal{C}} J(u,{v})}
\end{equation} 
where $v$ is set to the vector of ones $1_{K^2(N+1)}$, which means that no influence/opinion control is allowed;\\ 
\begin{equation}\label{eq-POAv}
\mathrm{PoA}_{v} = \frac{J(0_{K^2}, v^\star(0_{K^2}) )}{\min\limits_{(u,{v})\in\mathcal{C}} J(u,{v})}.
\end{equation} 
where $u$ is set to the vector of zeros $0_{K^2}$, which means that no epidemic control is allowed. For example, if the PoA equals 2 when the cost function is taken to be the global reproduction number only, it means that decentralization leads to a reproduction number twice higher, which is very significant since the epidemics spreads exponentially in the reproduction number.


Computing the above quantities relies on being able to globally minimize the sum-cost $J$. It is known that the sum-cost minimization problem is generically hard. For the game under consideration, it is possible to exploit the auxiliary game to dramatically decrease the computational complexity associated with the global minimization of $J$. This is what is stated through the next proposition.

\begin{prop}\label{Prop:Efficiency}
The global minimum of the sum-cost function $J$ can be found by solving a convex optimization problem.\hfill$\Box$
\end{prop}
\noindent\emph{$\mathbf{Proof.}$} 
See Appendix-C. \hfill$\blacksquare$

In the next subsection, we tackle the computation problem of the GNE of $\mathcal{G}$. 

\subsection{GNE determination algorithm}\label{subsec:determination}

In Sec.~\ref{subsec:existence_uniqueness}, we have shown that the game $\mathcal{G}$ has a unique $\mathrm{GNE}$. Here, we propose an algorithm to find this unique equilibrium point. To compute the GNE of $\mathcal{G}$ we again resort to the auxiliary game $\widetilde{\mathcal{G}}$ for which the GNE is much easier to compute. Indeed, one of the key ingredients of the algorithm is to use a gradient-type updating rule for minimizing $\widetilde{J}_k$, which is relevant since the auxiliary game is convex in the sense of Rosen \cite{rosen1965existence}. The function $\widetilde{J}_k$ is not only individually convex (i.e., w.r.t. $(u_k,v_k)$) but also jointly convex (i.e., w.r.t. $(u_,v)$), which is exploited to exhibit a Lyapunov function for the convergence analysis of the proposed algorithm. Using the notations introduced in Appendix-A, we denote by $\xi=(\xi_1,\ldots,\xi_K,\xi_{K+1})$ such that $\forall k\in\mathcal{K}$, $\xi_k=(\xi_{y_k},\xi_{\omega_k})$, where: $\xi_{y_k}=(\xi_{y_{k1}},\ldots,\xi_{y_{kK}})$ such that $\xi_{y_{k\ell}}\in\R$;
$\xi_{\omega_k}:=(\xi_{\omega_{k}}(0),\ldots,\xi_{\omega_k}(N))$ where $\forall n\in\{0,\ldots,N\}$, $\xi_{\omega_k}(n):=(\xi_{\omega_{k1,\ldots,1}}(n),\xi_{\omega_{k1,\ldots2}}(n),\ldots,\xi_{\omega_{kK\ldots K}}(n))$ such that $\forall (\ell_n,\ldots,\ell_0)\in\mathcal{K}^{(n+1)}$, $\xi_{\omega_{k \ell_n \ell_{n-1} \ldots \ell_0}}(n)\in\R$. 
For $k=K+1$ we denote by $\xi_{K+1}=(\xi_\lambda,\xi_x)$ where: $\xi_{\lambda}=(\xi_{\lambda}(0),\ldots,\xi_{\lambda}(N+1))$  such that $\forall n,\ \xi_{\lambda}(n)\in\R$ 
and $\xi_x=(\xi_x(0),\ldots,\xi_x(N+1))$ such that $\forall \ell\in\mathcal{K}$ and $\forall n$, the $\ell^\text{th}$-component of $\xi_x(n)$ is given by $\xi_{x_{\ell}}(n)\in\R$. 

Since the proposed algorithm is an iterative procedure, a natural question is whether the algorithm converges and to which convergence point. The following proposition provides the corresponding result.

\begin{prop}\label{Prop:Determination}
The Generalized Nash equilibrium seeking algorithm given in Tab.~\ref{tab:algo} converges to the $\mathrm{GNE}$ of $\mathcal{G}$.\hfill$\Box$
\end{prop}
\noindent\emph{$\mathbf{Proof.}$} See  Appendix-D.
\color{black}\hfill$\blacksquare$

\begin{algorithm}
\caption{Generalized Nash equilibrium seeking algorithm for $\mathcal{G}$}\label{tab:algo}
\begin{algorithmic}
\State\hspace{-1em}$\textbf{Initialization}:$  $t=0$,\\
         $\forall(k,\ell)\in\mathcal{K}^2$, $\forall n\in\{0,\ldots,N\}$, $\forall (\ell_n,\ldots,\ell_0)\in\mathcal{K}^{n+1}$,\\ 
         $\xi_{y_{k\ell}}^{(0)}\in\mathcal{Y}_{k\ell}$,  $\xi_{\omega_{k \ell_n \ell_{n-1} \ldots \ell_0}}^{(0)}(n) \in\mathcal{W}_{k \ell_n \ell_{n-1} \ldots \ell_0}(n)$\\
         $\xi_{\lambda}^{(0)}(n) \in\boldsymbol{\Lambda}$, $\xi_{x}^{(0)}(n)  \in\mathcal{X}$.\\
         Let $\overline{\delta}=(\overline{\delta}_1,\ldots,\overline{\delta}_K,\overline{\delta}_{K+1})>0$.\\
\State\hspace{-1em}$\textbf{Process}:$ $\forall k\in\mathcal{K}\cup\{K+1\}$,\\
         \[\displaystyle\frac{\mathrm{d}\xi_k}{\mathrm{d}t}=-\overline{\delta}_k\nabla_{\xi_k}\widetilde{J}_k+\hspace{-3.5em}\sum\limits_{j\in\{1\leq i\leq M:\ \widetilde{h}_i(\xi)>0\}}\hspace{-3em} \overline{\mu}_j\nabla_{\xi_k}\widetilde{h}_j(\xi),\]
         where:\\
         $\forall k\in\mathcal{K}\cup\{K+1\}$, $\widetilde{J}_k$ is given in \eqref{eq:tildeJk};\\
         $\widetilde{h}$ is given in \eqref{eq:tildeC}, $M=\mathrm{dim}(\widetilde{h})$ and $\widetilde{M}=\mathrm{dim}(\xi)$\\
        $\forall j\in\{1,\ldots,M\}$, $\widetilde{h}_j$ is the $j^\text{th}$-component of $\widetilde{h}$;\\
        $\overline{\mu}_j$ is the $j^\text{th}$-nonzeros element of $\overline{\mu}\in \R_{\leq 0}^{\overline{M}}$ where $\overline{M}\leq M $\\
        and $\overline{\mu}(\xi)=\left[\boldsymbol{\overline{H}}(\xi)^\top \boldsymbol{\overline{H}}(\xi)\right]^{-1}\boldsymbol{\overline{H}}(\xi)^\top g(\xi,\overline{\delta})\leq 0$\\
        where the matrix $\boldsymbol{\overline{H}}(\xi)\in\R^{\widetilde{M}\times \overline{M}}$ is composed by\\
        $\overline{M}\leq M $ linearly independent columns of\\ $\boldsymbol{H}(\xi)=[\nabla_{\xi}\widetilde{h}_{1}(\xi),\nabla_{\xi}\widetilde{h}_{2}(\xi),\ldots,\nabla_{\xi}\widetilde{h}_{M}(\xi)]$\\
        selected from $\nabla_{\xi} \widetilde{h}_j(\xi)$\\
        for $j\in\{i\in\{1,\ldots,M\} :\ \widetilde{h}_i(\xi)>0\}$.\\
\State\hspace{-1em}$\textbf{Output}:$  $\forall (k,\ell)\in\mathcal{K}^2$ and $\forall n\in\{0,\ldots,N\}$,\\
        \[u_{k\ell}^{\star}=\beta_{k\ell}^0-\lim\limits_{t\to+\infty}\exp(\xi_{y_{k\ell}}^{(t)})\]
        \[v_{k\ell}^{\star}(n) = \lim\limits_{t\to+\infty}\Big[\exp(\xi_{\omega_{k\ell k\ldots k}}^{(t)}(n))-\exp(\xi_{\omega_{\ell k \ldots k}}^{(t)} (n-1) )\Big].\]
\end{algorithmic}
\end{algorithm}


\section{Numerical performance analysis}\label{sec:Numerical-Analysis}


The goal of this section is to assess numerically the efficiency measures introduced in Section~\ref{subsec:EffectivenessofGNE}. The numerical analysis is conducted for COVID-19-type scenarios but \textbf{the proposed methodology may be applied to other types of epidemics including viral marketing-type ones.} To choose the parameters of the epidemic model, we have in part exploited the studies on Covid-19 that have been conducted in \cite{lasaulce2020efficient,casella2020can,salje2020estimating}. We assume a territory that is divided into $K=10$ geographical regions; the time horizon of the considered epidemic phase is set to $T= 40\ \mathrm{days}$ and regions apply $3$ awareness/influence campaigns at $ t_1=10\ \mathrm{days},\ t_2= 20\ \mathrm{days},\ t_3=30\ \mathrm{days}$. For simplicity we assume that $\forall k\in\mathcal{K}$ and $n\in\{1,2,3\}$,  $\theta_k^{\max}=\mu_{k}=\widehat{\psi}_{k}(n)=\mathrm{R}_k^{\max}=+\infty$. For the epidemic model parameters, it is assumed that: $\forall k\in\mathcal{K}$, $\gamma_k=0.2$. \BLUE{When the in-degree per region in the social network equals $0$, we take $\forall n,\ \boldsymbol{P}(n)=\boldsymbol{I}_K$; in the other cases $\forall n,$ \[p_{k\ell}(n)=\left\{\begin{array}{ll}
  1/\mbox{degree of $k$} &\text{ if }k \text{ and }\ell\text{ are connected}\\
  &\hspace{0.5cm}\text{in the social network,}\\
  0& \text{otherwise.}
   \end{array}\right.\]} The perturbation matrix $\widehat{\boldsymbol{B}}$ is given by $\widehat{\boldsymbol{B}}=0.5\boldsymbol{B}^0$ where  $\boldsymbol{B}^0=\boldsymbol{B}\odot\boldsymbol{\widetilde{A}}$ and $\boldsymbol{B}=$
   
   \[{\tiny\begin{pmatrix}
0.37&0.03&0.06&0.01&0.02&0.02&0.01&0.08&0.01&0.08\\
0.05&  1.00 &0.08&0.22&0.15&0.25&0.27&0.19&0.05&0.26\\
0.07&0.14&1.00&0.13&0.14&0.08&0.05&0.04&0.15&0.07\\
0.22&0.21&0.01 &0.88&0.05&0.23&0.14&0.01&0.16&0.21\\
0.01&0.11&0.20&0.09&0.72&0.18&0.10&0.18&0.15&0.11\\
0.21&0.17&0.06&0.08&0.06&0.90&0.19&0.23&0.17&0.16\\
0.02&0.06&0.05&0.07&0.07&0.07&0.24&0.08&0.02&0.03\\
0.15&0.10&0.22&0.26&0.01&0.13&0.03&1.00&0.15&0.13\\
0.04&0.01&0.01&0.04&0.01&0.01&0.01&0.05&0.21&0.01\\
0.06&0.03&0.05&0.05&0.01&0.07&0.03&0.06&0.06&0.29\\[1mm]
\end{pmatrix},}\]
where $\boldsymbol{B}$ is diagonal dominant.
\[\text{and }[\boldsymbol{\widetilde{A}}]_{k\ell}=\left\{\begin{array}{ll}
  1 &\text{ if }k \text{ and }\ell\text{ are connected}\\
  &\hspace{0.5cm}\text{in the epidemic graph,}\\
  10^{-10}& \text{otherwise.}
   \end{array}\right.\]\vspace{-0.1em}
   
  The initial state is given by $s(0)=1-i(0)$,  \[\begin{array}{l}
\hspace{-0.25em}i(0)\hspace{-0.25em}=10^{-2}\cdot (0.2,0.1,2,0.1,3,0.5,2.5,1,2,0.1),\\
\hspace{-0.25em}\theta(0)\hspace{-0.25em}=\hspace{-0.25em}(0.59, 0.25 , 0.25 , 0.46 , 0.26 , 0.68 , 0.16 , 0.24 , 0.71 , 0.6).
\end{array}\]  
\BLUE{To simplify the numerical analysis, throughout this section we consider $\forall k\ b_k^{\mathrm{local}}=d_k=0$. This choice simplifies the individual costs allowing us to highlight the trade-off between the epidemic management and the opinion dynamics control.}\\
\textit{Influence of the epidemic and influence graphs on the PoA.}
\begin{table}[H]
        \centering
  \begin{tabular}{l|c|c|c|c|c|}
      $\mathrm{PoA}_{uv}$ \multirow{4}{*}{\vspace*{-5em}\rot{\text{\small{Epidemic graph}}}}  &\multicolumn{4}{c}{\hspace{4em}\text{\small{Influence graph}}} \\
                 \cline{1-6}

           &Degree&0& 2& 6 & 10\\
         \cline{2-6}

        &0   
         & $1.34$  & $1.38$ & $1.30$ & $1.23$\\       

                 \cline{2-6}
      & 2
          & $1.46$  & $1.60$ & $1.67$ &  $1.66$\\
                 \cline{2-6}
       &6
         &$1.67$  & $1.76$  & $1.83$ & $1.81$\\
                          \cline{2-6}
         &10
          & $1.78$  & $1.85$ & $1.92$ & $1.91$\\
                 \cline{2-6}               
    \end{tabular}                        

\vspace{0.2cm}
\caption{\BLUE{The table shows that the denser the epidemic graph the higher the PoA. For the opinion influence graph, the PoA is not necessarily the highest for the densest graph.}}
    \label{tab:simulation}
\end{table}
As expected the PoA is more sensitive to the interactions density in the epidemic graph than in the opinion influence one.
In Tab.~\ref{tab:simulation} we set $a_k=c_k=1$, $b_k^{\mathrm{global}}=10$ and present the PoA ($\mathrm{PoA}_{uv}$) for different values for the degrees of the two graphs. The PoA is averaged over a total of 1600 realizations of the epidemic and social graphs. The simulation results show that the largest value for $\mathrm{PoA}_{uv}$ is $1.92$ and is achieved when all regions are interconnected both for the epidemic graph and influence graph. The smallest value for $\mathrm{PoA}_{uv}$ value is $1.23$ and is obtained when there is no interconnection in the epidemic graph and when the influence graph is fully connected. The study also reveals that there is no correlation between the average degree per agent in the influence graph and $\mathrm{PoA}_{uv}$. However, an increase in the degree per agent in the epidemic graph results in an increase in $\mathrm{PoA}_{uv}$. Notably, even when the epidemic graph and social network are not interconnected, the $\mathrm{PoA}_{uv}$ value is still greater than one, indicating the presence of efficiency loss. The study highlights the importance of considering both the epidemic graph and the influence graph for designing decentralized decision-making processes for managing epidemics.\\
 
\begin{figure}[h]
\centering
\includegraphics[width=0.8\linewidth]{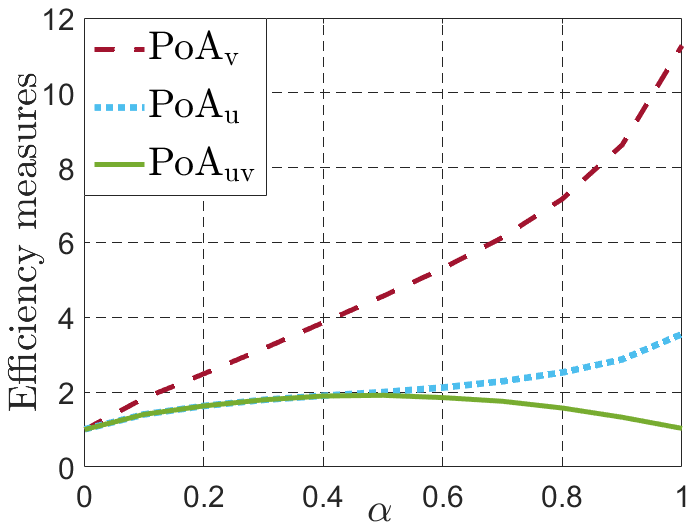}
\caption{\label{fig1}{{\BLUE{Bottom curve: When each region controls the epidemic both through physical measures ($u$) and opinion ($v$), the maximum value reached for the PoA is $2$, which is already significant. Middle curve: When only physical measures are controlled and the opinion is left to evolve freely, the PoA can be as large as $3.6$, showing the loss of non-controlling the opinion. Top curve: when each region only controls its opinion, very large values for the PoA can be reached ($>10$), showing the irrelevance for decentralized management when based only on opinion control.} 
}}}

\end{figure}

\begin{figure}[h]
\vspace{0.5cm}
\includegraphics[width=0.8\linewidth]{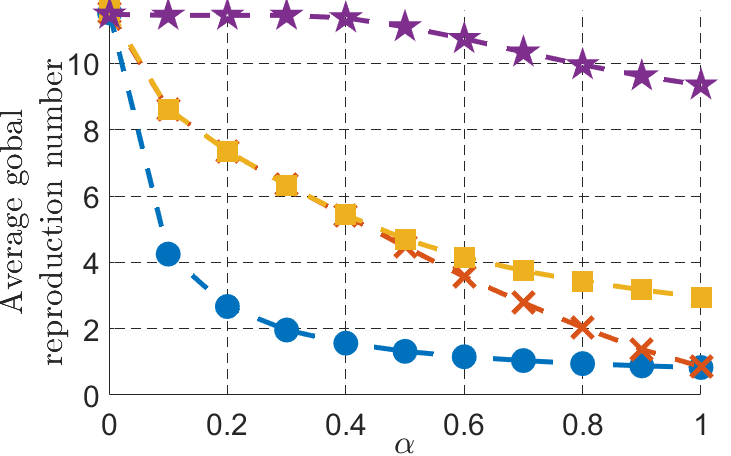}
\includegraphics[width=0.9\linewidth]{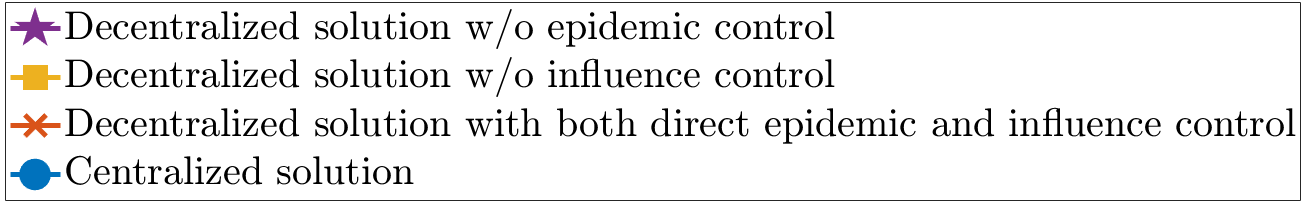}
\caption*{(a) Plot of $\displaystyle{ \sum_{n=0}^N\frac{\rho\left(\boldsymbol{{D}}_{\gamma}^{-1} \left(\boldsymbol{{B}^0}-\boldsymbol{U}+\mathrm{Diag}(\theta(n))\widehat{\boldsymbol{B}}\right)\right)}{N+1}}$ vs $\alpha$.}
\centering 
\vspace{0.5cm}
\includegraphics[width=0.8\linewidth]{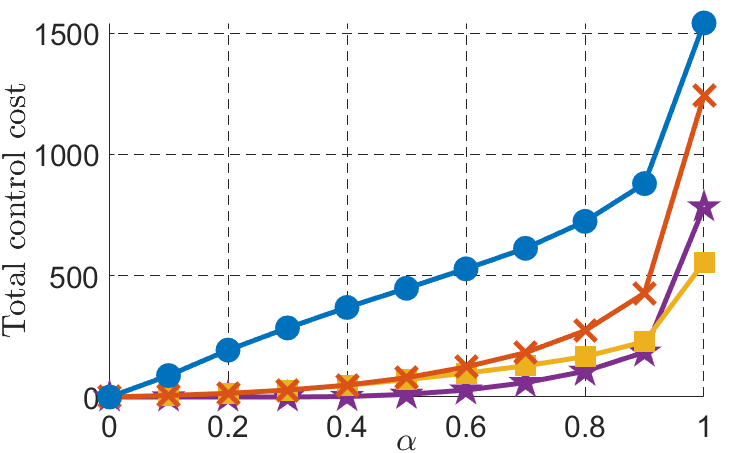}
\includegraphics[width=0.9\linewidth]{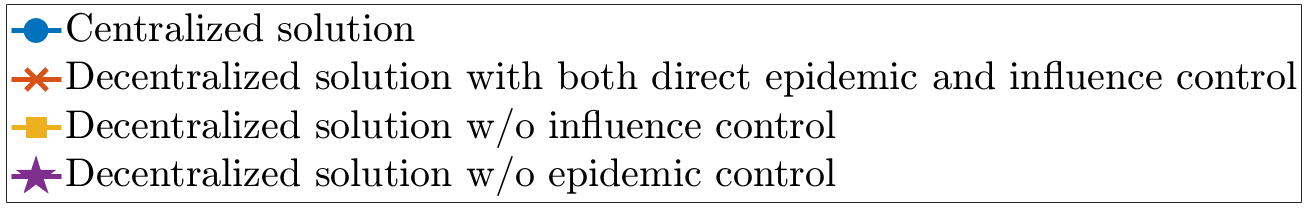}
\caption*{(b) $\displaystyle\sum_{k=1}^K\Big[-\sum_{\ell\in\mathcal{N}_k}\log(\frac{\beta_{k\ell}^0-u_{k\ell}}{\beta_{k\ell}^0})+\sum_{n=0}^N$ $\displaystyle-\sum_{\ell\in\mathcal{N}_k^{\mathrm{S},n}}\log(v_{k\ell}(n))\Big]$  vs. $\alpha$.}
\vspace{0.2cm}
\caption{\label{fig2}{{\BLUE{One of the key information the above figures provide is the loss in terms of cost function induced by decentralization for a given global reproduction number. For instance, a target reproduction number of $2$ is reached with a centralized management for $\alpha=0.3$ whereas it is reached with $\alpha=0.8$ for decentralized management. This difference in terms of $\alpha$ can be translated in terms of economic cost (e.g., in billions of US dollars) by using existing quantitative analyses \cite{lasaulce2020efficient}.}}}}
\end{figure}

\begin{figure}[h]
\centering
\vspace{0.5cm}
\includegraphics[width=0.8\linewidth]{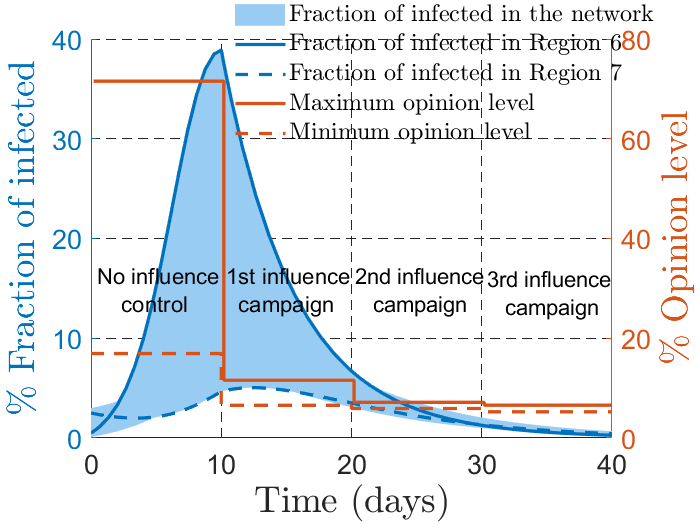}
\caption{\label{fig3}{Evolution of the fractions of infected and opinion levels for the different regions. The effect of influence campaigns on the fractions of infected appears very clearly.}}
\end{figure}

\begin{figure}[h]
\centering
\vspace{0.5cm}
\includegraphics[width=0.8\linewidth]{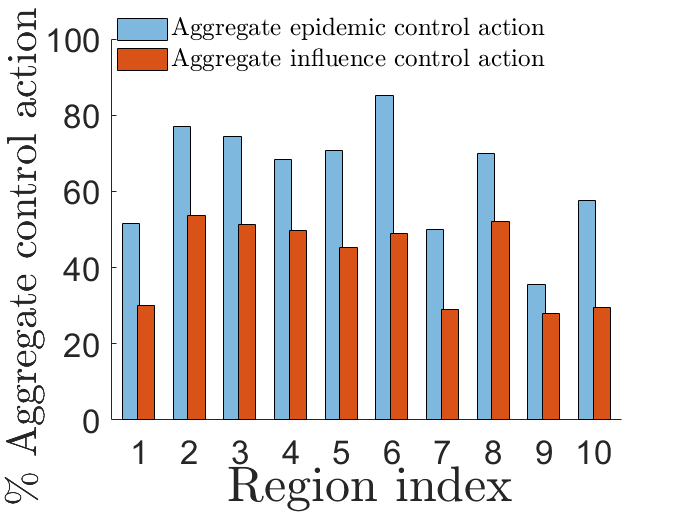}
\caption{\label{fig4}{{The figure provides the control action intensity for the different regions. The corresponding values have to be put in correlation with the local situation of the epidemic, which is in part related to the values of the natural reproduction numbers.}}}
\end{figure}

\textit{Influence of the cost function and control actions on the PoA.}\\ The cost function $J_k$ comprises a collective term (namely, the term weighted by $b_k^{\mathrm{global}}$) which is common to all the players whereas all the other terms are individual terms. To study the impact of the collective and individual terms on the PoA, we introduce the parameter $\alpha \in [0,1]$ which is used for Fig.\ref{fig1} and Fig.\ref{fig2} and is defined as follows: $a_k=1-\alpha,\ b_k^{\mathrm{global}}=10\times\alpha$ and $c_k=1-\alpha$  for all $k\in\mathcal{K}$. Additionally, for Fig.\ref{fig1} to Fig.\ref{fig4}, we will assume $\boldsymbol{{B}}^0=\boldsymbol{{B}}$ and $\forall n,\ [\boldsymbol{P}(n)]_{k\ell}=1/10$. Fig.~\ref{fig1} represents the different efficiency measures ($\mathrm{PoA}_{uv}$ in \eqref{eq-POAuv}, $\mathrm{PoA}_u$ in \eqref{eq-POAu} and $\mathrm{PoA}_v$ in \eqref{eq-POAv}) against $\alpha$. When $\alpha=1$, the game becomes a team game and the GNE coincides with a local minimum point of the common cost function $J_k= b_k^{\mathrm{global}}\displaystyle{{\sum_{n=0}^{N+1}}\rho\left(\boldsymbol{{D}}_{\gamma}^{-1} \left(\boldsymbol{{B}^0}-\boldsymbol{U}+\mathrm{Diag}(\theta(n))\widehat{\boldsymbol{B}}\right)\right)}$; the fact that $\mathrm{PoA}_{uv} = 1$ indicates the local minimum coincides with the global minimum and decentralization induces zero optimality loss. When the opinion is not controlled, this result is no longer true since the PoA is as large as almost $4$, which is very significant. If the epidemic is not controlled and only the influence is controlled, the PoA reaches values as large as $11$. Now, when both epidemic and influence are controlled, the largest value for the PoA is $\mathrm{PoA}_{uv} \sim 2$, which is reached when $\alpha=0.5$; this corresponding efficiency loss is still significant.

Fig.~\ref{fig2} depicts the average global reproduction number (Fig.~\ref{fig2}.a) and total control cost (Fig.~\ref{fig2}.b) against $\alpha$ for four distinct control strategies: the GNE strategy defined in \eqref{eq-POAuv}); the GNE strategy with no influence control (the strategy profile $(u^\star(1_{K^2(N+1)}),1_{K^2(N+1)})$ defined in \eqref{eq-POAu}); the GNE strategy with no epidemic control (the strategy profile $(0_{K^2},v^\star(0_{K^2}))$ defined in \eqref{eq-POAv}); and the optimal centralized strategy (the strategy profile that minimizes $\argmin\limits_{(u,v)\in\mathcal{C}}J(u,v)$). The figure provides several insights. For example, one sees the impact on the global reproduction number of the fact that regions care about their individual socio-economic cost. For example, for $\alpha \sim 0.8$, a centralized solution would yield a value of less than $1$ for the reproduction number whereas it reaches $2$ for a decentralized management. If the opinion cannot be controlled, then this value becomes about $3.7$, showing the importance of opinion influence. Now, when the epidemic cannot be directly controlled (namely, through $u$), the impact of the opinion influence becomes negligible and the reproduction number reaches values as large as $9.3$ to $11.4$. Fig.~\ref{fig2}.b illustrates well the effect of decentralization and control actions on the total control cost.\\ 


\textit{Analysis of the control actions.}\\ For the preceding simulation results, the focus has been on the effect of decentralization and control actions on global epidemic management efficiency. Here, we want to have more insights into the equilibrium control actions themselves both in space (over the regions) and time. For this, we define the aggregate GNE control action in $\%$ as follows:  $\overline{u}_k^\star =   \frac{100}{K}\times\sum_{\ell=1}^K \frac{u_{k\ell}^\star}{\beta_{k\ell}^0}$; $\overline{v}_k^\star(t_1)=\frac{100}{K}\times \sum_{\ell=1}^K (1-v_{k\ell}(t_1))$. In Fig.~\ref{fig3} to Fig.~\ref{fig4}, we set $a_k=0.1$,  $b_k^{\mathrm{global}}=9$ and $c_k=0.1$ for all $k\in\mathcal{K}$.  Fig.~\ref{fig4} depicts the value of the epidemic and opinion aggregate control actions for the $10$ regions. These values have to be put in correlation with the parameters of the epidemics and, in particular, with the natural reproduction numbers namely, the elements of the diagonal of the $\boldsymbol{B}$ matrix: $(0.37, 1, 1, 0.88, 0.72, 0.9, 0.24, 1, 0.21, 0.29))$. The intuition that regions having a higher natural reproduction number should undergo more severe measures is confirmed. But the proposed methodology says more than that since it has also the advantage of quantifying this relationship and thus providing the severity level each region should apply. Now, we look at the time aspect. In Fig.~\ref{fig3}, we represent the evolution of the fractions of infected and the opinions of the regions. For the sake of clarity, we represent the proportion of infected in each region by a blue shape rather than plotting 10 curves. The direct epidemic control actions and the influence control actions are fixed at the $\mathrm{GNE}$ strategy for the whole time period ($40$ days) by
 $\boldsymbol{U}^\star=10^{-2}\times$
 
 {\tiny\[\begin{pmatrix}
    36.3 & 2.8 & 5.6 & 0.5 & 2 & 2 & 0 & 8 & 0.5 & 7.5\\
    4 & 99.4 & 7.4 & 21.4 & 15 & 25 & 26.2 & 18.8 & 4.4 & 24.9\\
     6.6 & 14 & 99.4 & 13.3 & 13.7&8&4.5 & 4 & 13.6 & 6.3\\
    21 & 20.8  & 0.9&  88 &  5 & 22.7 & 13.3  & 0.1 & 15 & 20.6\\
    0  & 11.0 & 19.7  & 8.6 & 72 & 17.8 &  9.2 & 17.8 & 13.7  & 9.5\\
    20 & 16.7  & 5.5  & 8  & 6 & 90&  18.4 & 22.7 & 16.0 & 16\\
    1.1 &  6.4&   4.7  & 6.6 &  6.5 &  6.7 & 23.6  & 0.2&   1.2 &  2.4\\
    14 &  9.8 & 22 & 25.7  & 0.5 & 13 &  2.7&  99.4&  14.4&  12.8\\
    3.8 &  0.9 &  0.5 &  4.5  & 1.4 &  0.2 &  0.5 &  5&  20.1 &  1\\
    5.2 &  2.9 &  4.6 &  4.9 &  0.6 &  7 &  2.1 &  6&  5&  28.4\\[1mm]
\end{pmatrix},\]}

$\boldsymbol{V}^\star(1)=10^{-2}\times$
{\tiny \[\begin{pmatrix}   14&32&32&18&31&12&49&33&11&13\\
 8&20&20&11&19& 7&31&21& 7& 8\\
 9&21&21&12&20& 8&32&22& 7& 9\\
 9&22&22&12&21& 8&33&22& 7& 9\\
10&24&24&13&23& 9&37&25& 8&10\\
 9&22&22&12&22& 8&34&23& 8& 9\\
15&35&35&19&34&13&56&36&12&15\\
 9&20&20&11&20& 7&31&21& 7& 8\\
18&43&43&23&41&15&99&44&15&17\\
14&34&33&18&32&12&53&35&12&14\\[1mm]
\end{pmatrix},\]}
$\boldsymbol{V}^\star(2)\approx 1_{K}\times 1_{K}^\top$ and $\boldsymbol{V}^\star(3)=1_{K}\times 1_{K}^\top$. The effect of the opinion influence is obvious. It is seen that the fractions of infected decrease significantly after only one influence campaign; for example, the fraction of infected in the most infected region decreases from $40\%$ to less than $10\%$. The impact of the subsequent campaigns is still positive but much less significant.


\section{Summary and conclusions}

In this paper, we propose a methodology to assess the effects of decentralization of the management of an epidemic \BLUE{in presence of an opinion dynamics.} For this, a game model \BLUE{which implements a good tradeoff between realism (which is here to implement features of practical interest) and analytical tractability (e.g., to conduct the chosen solution concept analysis) is proposed. The presence of the coupled constraint (namely, the last constraint of (4)) 
has led us to a solution concept that is more involved than the NE, that is the GNE. The GNE analysis (existence, uniqueness, determination) is seen to be non-trivial but can be made possible by 
exploiting an appropriate auxiliary game. The corresponding analysis is not only useful in itself and to support the proposed modeling but also constitutes a required preliminary work to be able to assess the efficiency loss induced by decentralization; in particular, uniqueness can be proved thanks to the convexity properties of the auxiliary game and the sum-cost minimization problem is shown to be a convex problem as well. To assess global efficiency, the PoA is retained; although the PoA is known to be difficult to express or bounded in general, it is a widely used metric for game-theoretic analyses \cite{papadimitriou2001algorithms}\cite{roughgarden2002bad}\cite{lasaulce2011game}. The conducted numerical analysis allows one to provide numerous insights on the problem of decentralized epidemic management, which can be exploited in practice and serve as a basis to elaborate more realistic or complex models. We would like to emphasize the following take-away messages: 1. The nature of the epidemic and opinion graphs impact the PoA in a non-trivial way, which would need to be formalized in a separate work. Our results tend to indicate that the PoA increases with the degree of the epidemic graph. The influence of the opinion graph nature seems a more involving issue; 2. For typical simulation settings \cite{lasaulce2020efficient}, the PoA can reach $2$ even if each region locally controls both the virus propagation and the opinion. When the region cost is dominated by the global reproduction number, a PoA of $2$ means that the decentralized management leads to a spatially averaged reproduction number that is $2$ times larger than the centralized scenario, which is a huge difference in terms of propagation (the number of infected being exponential in the reproduction number); 3. If the opinion is not controlled, the PoA can reach values as large as $3.6$ and when the epidemic is only controlled through opinion, the PoA blows up and reaches values larger than $10$, showing the irrelevance of decentralization by just relying on influence management; 4. Our analysis constitutes a first step to quantify the economic losses induced by decentralization for a given target in terms of global reproduction number. This can done by using quantitative analyses such as \cite{lasaulce2020efficient}.} All these very encouraging results suggest extensions of the proposed model. The relevance of a dynamic game model might be studied. New constraints might be added such as constraints on the fractions of infected. Scalability issues might be analyzed by treating the case of a large number of regions and possibly handled by the use of a mean-field game approach. Also, the developed approach might be combined with a data-oriented approach.

\section*{Appendix}
\subsection{Auxiliary Game}\label{subsec:appendix-auxiliarygame}
\noindent In view of the dynamic of $\theta$, $\forall k,n$ the drift $\theta_k(n)$ is a posynomial function w.r.t. the awareness action $v$ i.e., $ \displaystyle \theta_{k}(n+1)= \sum_{\ell_n\in\widehat{\mathcal{N}}_k(n)}\sum_{\ell_{n-1}\in\widehat{\mathcal{N}}_{\ell_n}(n-1)}\ldots\sum_{\ell_0\in\widehat{\mathcal{N}}_{\ell_0}(0)}\alpha_{k \ell_n \ell_{n-1} \ldots \ell_0}(n)\times$\newline $ v_{k\ell_n}(n)v_{\ell_n,\ell_{n-1}}(n-1)\ldots v_{\ell_1,\ell_0}(0)\theta_{\ell_0}(0)$, where\hspace{1em}
 
 \noindent $\alpha_{k,\ell_n,\ell_{{n-1}},\ldots,\ell_0}(n)\geq 0$. In the sequel, we denote by: $\forall n\in\{0,\ldots,N\}$ and $\forall (k,\ell_n,\ldots,\ell_0)\in\mathcal{K}^{n+2}$, $\omega_{k \ell_n \ell_{n-1} \ldots \ell_0}(n)=v_{k\ell_n}(n)v_{\ell_n,\ell_{n-1}}(n-1)\ldots v_{\ell_1,\ell_0}(0)$. Let $\varepsilon_x>0$ sufficiently small such that, from Assumption \ref{assumption} and the theory of nonnegative matrix in \cite{meyer2000matrix}, we use Perron-Frobenius lemma. For all $(u,v)\in\mathcal{U}\times\mathcal{V}$ and $n\in\{0,\ldots,N\}$ we have: $\rho\left(\boldsymbol{D}_{\gamma}^{-1} \left(\boldsymbol{{B}^0}-\boldsymbol{U}+\mathrm{diag}(\theta(n))\boldsymbol{\widehat{B}}\right)\right) =\min
 \lambda(n)$
\begin{eqnarray}
          &&\hspace{-.5cm}\text{s.t.}\ \exists\ x(n)\  \text{with} \quad \sum_{\ell=1}^K x_{\ell}(n)\leq 1,\quad x_{\ell}(n)>\varepsilon_x\label{eq:rho}\\
          && \hspace{-1cm}\text{and}\quad\boldsymbol{D}_{\gamma}^{-1} \left(\boldsymbol{{B}^0}-\boldsymbol{U}+\mathrm{diag}(\theta(n))\boldsymbol{\widehat{B}}\right)x(n)\leq \lambda(n)x(n).\notag
\end{eqnarray}
Let us consider the following changes of variables such that $(k,\ell,\ell_n,\ldots,\ell_{0})\in\mathcal{K}^{n+3}$, 
\begin{equation}\label{eq:uV}\begin{array}{l}
     \hspace{-1em}\xi_{\omega_{k \ell_n \ell_{n-1} \ldots \ell_0}}(n)=\log{(\omega_{k \ell_n \ell_{n-1} \ldots \ell_0}(n))}, \\   \hspace{-1em}\xi_{y_{k\ell}}=\log{(\beta_{k\ell}^0-u_{k\ell})}.
     \end{array}\end{equation}In what follows, we define the operator $\mathrm{Exp}(\cdot)$ which corresponds to the component-wise exponential operator. Let $\rho^{\min}:=\rho(\boldsymbol{D}_{\gamma}^{-1}(\boldsymbol{B^0}-\boldsymbol{U}^{\max}))$, $\rho^{\max}:=\rho(\boldsymbol{D}_{\gamma}^{-1}(\boldsymbol{B^0}-\boldsymbol{U}^{\min})+\boldsymbol{\widehat{B}})$, where $\boldsymbol{U}^{\min,\max}=[u_{k\ell}^{\min,\max}]_{1\leq k,\ell\leq K}$. For all $k\in\mathcal{K}$, we denote the action profile of the $k^{\text{th}}$ auxiliary player by $\xi_k:=(\xi_{y_k},\xi_{\omega_k})$, where: $\xi_{y_k}:=(\xi_{y_{k,1}},\ldots,\xi_{y_{k,K}})$ and $\forall \ell\in\mathcal{K},\ \xi_{y_{k\ell}}\in\mathcal{Y}_{k\ell}:=[\log(\beta_{k\ell}^0-u_{k\ell}^{\max}),\log(\beta_{k\ell}^0-u_{k\ell}^{\min})]$; $\xi_{\omega_k}:=(\xi_{\omega_{k}}(0),\ldots,\xi_{\omega_k}(n))$ and $\forall n\in\{0,\ldots,N\}$, $\xi_{\omega_k}(n):=(\xi_{\omega_{k,1,\ldots,1}}(n),\xi_{\omega_{k,1,\ldots,2}}(n),\ldots,\xi_{\omega_{k,K,\ldots,K}}(n))$ such that $\forall (\ell_n,\ldots,\ell_0)\in\mathcal{K}^{n+1}$, $\xi_{\omega_{k \ell_n \ell_{n-1} \ldots \ell_0}}(n)\in\mathcal{W}_{k \ell_n \ell_{n-1} \ldots \ell_0}:=[\log(v_{k,\ell_n}^{\min})+\ldots,\log(v_{\ell_1,\ell_0}^{\min}),\log(v_{k,\ell_n}^{\max})+\ldots,\log(v_{\ell_1,\ell_0}^{\max})]$.
We consider an additional player of index $K+1$ with the corresponding action profile $\xi_{K+1}:=(\xi_\lambda,\xi_x)$ where: $\xi_{\lambda}:=(\xi_{\lambda}(0),\ldots,\xi_{\lambda}(N+1))$ such that $\forall n\in\{0,\ldots,N+1\}$, $\xi_{\lambda}(n)\in\mathbf{\Lambda}:=[\log(\rho^{\min}),\log(\rho^{\max})]$; $\xi_x:=(\xi_x(0),\ldots,\xi_x(N+1))$ such that $\forall n\in\{0,\ldots,N+1\}$, $\xi_x(n)\in\mathcal{X}:=[\log(\varepsilon_x),0]^K$. In the following, we denote the complete auxiliary action profile by $\xi:=(\xi_1,\ldots,\xi_K,\xi_{K+1})$ and for all $ k\in\mathcal{K}$ and $n\in\{0,\ldots,N\}$,\\
$\displaystyle \widetilde{\theta}_k(n+1)=\sum_{\ell_n\in\widehat{\mathcal{N}}_k(n)}\sum_{\ell_{n-1}\in\widehat{\mathcal{N}}_{\ell_n}(n-1)}\ldots\sum_{\ell_0\in\widehat{\mathcal{N}}_{\ell_1}(0)}\Big[$
\begin{equation}\label{eq:tildethetak}
   \alpha_{k \ell_n \ell_{n-1} \ldots \ell_0}(n)\exp(\xi_{\omega_{k \ell_n \ell_{n-1} \ldots \ell_0}}(n))\theta_{\ell_0}(0)\Big].
\end{equation}

The generalized form of the auxiliary static game under consideration is therefore given by:
\begin{equation}\label{eq-gameaux}
 \widetilde{\mathcal{G}}:=\Big(\mathcal{K}\cup\{K+1\}, \Big(\boldsymbol{\Pi}_k\widetilde{\mathcal{C}}\Big)_{1\leq k\leq K+1},\big(\widetilde{J}_k\big)_{1\leq k\leq K+1}\Big),
\end{equation} where the action spaces and utilities are as follows.
\begin{equation}\label{eq:tildeJk}
\begin{array}{l}
      \displaystyle\widetilde{J}_k(\xi):= - a_k  \sum_{\ell\in\mathcal{N}_k} [\xi_{y_{k\ell}}-\log(\beta_{k\ell}^0)]\\
 \displaystyle+b_k^{\mathrm{local}}\sum_{n=0}^{N+1}\sum_{\ell\in\mathcal{N}_k}\frac{\displaystyle\exp{(\xi_{y_{k\ell}})}+\widehat{\beta}_{k\ell}\widetilde{\theta}_k(n)}{\gamma_k}\\
      \displaystyle-c_k\sum_{n=1}^N\sum_{\ell\in\widehat{\mathcal{N}}_k(n)}\hspace{-0.5em}\Big[\xi_{\omega_{k\ell k\ldots k}}(n)-\xi_{\omega_{\ell \ldots k}}(n-1)\Big]\\
      \displaystyle-c_k\sum_{\ell\in\widehat{\mathcal{N}}_k(0)}\xi_{\omega_{k\ell}}(0)
      +d_k\displaystyle\sum_{n=0}^{N+1}\widetilde{\theta}_k(n+1).\\      
\displaystyle\widetilde{J}_{K+1}(\xi):=\sum_{n=0}^{N+1}\exp(\xi_{\lambda}(n)).
\end{array}\end{equation}
\noindent$\widetilde{\mathcal{C}}:=\Big\{\xi:\forall n\in\{0,\ldots,N\}$, $m\in\{0,\ldots,N+1\}$, $(k,\ell,\ell_n,...,\ell_0)\in\mathcal{K}^{n+3},$ $\log(\rho^{\min})\le\xi_{\lambda}(m)\leq\log(\rho^{\max}),$
     \begin{equation}\label{eq:tildeC}\begin{array}{l}
   \boldsymbol{D}_{\gamma}^{-1} \left(\mathrm{Exp}(\xi_{y})+\mathrm{diag}(\theta(m))\boldsymbol{\widehat{B}}\right)\mathrm{Exp}{(\xi_{x}(m))}\odot\\
     \hspace{8em}\Big[\mathrm{Exp}(-\xi_{\lambda}(m)1_K-\xi_{x}(m))\Big]\leq 1_K\\
     \log{(\beta_{k\ell}^0 -u_{k\ell}^{\max})}\le\xi_{y_{k\ell}}\leq\log{(\beta_{k\ell}^0-u_{k\ell}^{\min})}, \widetilde{\theta}_{k}(m)\leq \theta_k^{\max},\\
      -\xi_{\omega_{k \ell_n \ell_{n-1} \ldots \ell_0}}(n)\leq-(\log(v_{k\ell_n}^{\min})+\ldots+\log(v_{\ell_1\ell_0}^{\min}))\\
      \xi_{\omega_{k \ell_n \ell_{n-1} \ldots \ell_0}}(n)\leq\log(v_{k\ell_n}^{\max})+\ldots+\log(v_{\ell_1\ell_0}^{\max}),\\    
      -\xi_{x_{\ell}}(m)\leq -\log{(\varepsilon_x)},      \hspace{1em}\sum_{\ell=1}^K\exp{(\xi_{x_{\ell}}(m))}\leq 1\\
   \displaystyle \sum_{\ell\in\mathcal{N}_k}\left[\exp{(\xi_{y_{k\ell}})}+\widehat{\beta}_{k\ell}\widetilde{\theta}_k(m)\right]\Big/\gamma_k\leq\mathrm{R}_k^{\max},\\
   \displaystyle-a_k\sum_{\ell\in\mathcal{N}_k}[\xi_{y_{k\ell}}-\log(\beta_{k\ell}^0)]\leq \phi_{k},
        \end{array}  \end{equation}
        \begin{equation*}\begin{array}{l}
      \hspace{-1em}\displaystyle -c_k\sum_{\ell\in\widehat{\mathcal{N}}_{k}(n)}\hspace{-1em}\left[\xi_{\omega_{k\ell k\ldots k}}(n+1)-\xi_{\omega_{\ell k\ldots k}}(n))\right]\leq\widehat{\psi}_{k}(n+1),\\
     \hspace{-1em}\displaystyle -c_k\sum_{\ell\in\widehat{\mathcal{N}}_k(0)}\xi_{\omega_{k\ell}}(0)\leq \widehat{\psi}_{k}(0)\Big\}=:\{\xi:\widetilde{h}(\xi)\leq 0\},
     \end{array}
     \end{equation*}where $\widetilde{h}(\xi)$ is jointly convex w.r.t. $\xi$. The auxiliary players are denoted by index $k\in\{1,\ldots,K+1\}$ where the player $K+1$ is an additional player that we consider in our analysis; the action space for Player $k\in\mathcal{K}\cup\{K+1\}$ is given by $\boldsymbol{\Pi}_k\widetilde{\mathcal{C}}$ which is the projection of the sharing constraint set $\widetilde{\mathcal{C}}$ over the action profile of the auxiliary player $k$. It has to be noted that the game $\widetilde{\mathcal{G}}$ is a convex static and strategic game played in one shoot.

In this paper, we show that the properties of the $\mathrm{GNE}$ of $\widetilde{\mathcal{G}}$ coincide with those of the game $\mathcal{G}$. The Definition of the $\mathrm{GNE}$ for the game $\widetilde{\mathcal{G}}$ is characterized by what follows. We call $\xi^{\mathrm{GNE}}\in\widetilde{\mathcal{C}}$ a Generalized Nash equilibrium point of $\widetilde{\mathcal{G}}$ if $\forall k\in\mathcal{K}$, 
 \begin{equation}\label{eq:GNEgameaux}
   \xi_k^{\mathrm{GNE}}\in\argmin\limits_{\xi_k\in\boldsymbol{\Pi}_k \widetilde{\mathcal{C}}}\widetilde{J}_k(\xi_k,\xi_{-k}^{\mathrm{GNE}}).
   \end{equation}


\subsection{Proof of the Proposition \ref{Prop:Existence_Uniqueness}}

\noindent \emph{\textbf{\underline{Existence of a $\mathrm{GNE}$:}}}

 According to \cite[Thm. 3.1]{dutang2013existence}, the game $\widetilde{\mathcal{G}}$ has at least one GNE, $\xi^{\mathrm{GNE}}\in\widetilde{\mathcal{C}}$ since $\forall k\in\mathcal{K}$: $\boldsymbol{\Pi}_k\widetilde{\mathcal{C}}$ is nonempty, convex and compact subset of Euclidean space; $\widetilde{\mathcal{C}}$ is both upper-semicontinuous and lower-semicontinuous (e.g., \cite[Proposition 4.1-4.2]{dutang2013existence}); $\widetilde{\mathcal{C}}$ is nonempty, closed, convex; $\widetilde{J}_k$ is continuous in $\widetilde{\mathcal{C}}$ and $\forall \xi_{-k}\in\prod_{-k}\widetilde{\mathcal{C}}$, $\xi_k\mapsto \widetilde{J}_k(\xi_k,\xi_{-k})$ is quasiconvex on $\boldsymbol{\Pi}_k\widetilde{\mathcal{C}}$. 

Now, we will prove that the $\mathrm{GNE}$ strategies of $\mathcal{G}$ are given by those of $\widetilde{\mathcal{G}}$. Let $\xi^{\mathrm{GNE}}$ a Generalized Nash equilibrium of $\widetilde{\mathcal{G}}$ and we denote by ${u}^\star$ and $v^\star$ after the change of variable in \eqref{eq:uV}. In view of \eqref{eq:rho} it follows that: \[\widetilde{J}_{K+1}(\xi^{\mathrm{GNE}})=\displaystyle\sum_{n=0}^N\rho(\boldsymbol{\Gamma^{-1}}(\boldsymbol{B^0}-\boldsymbol{{U}^\star}+\mathrm{diag}({\theta}(n)^\star)\boldsymbol{\widehat{B}})),\]
and $\forall k\in\mathcal{K}$,
\[\begin{array}{ll}
&\displaystyle\widetilde{J}_k(\xi^{\mathrm{GNE}})=-a_k  \sum_{\ell\in\mathcal{N}_k} \log\left(\frac{\beta_{k\ell}^0-{u}_{k\ell}^\star}{\beta_{k\ell}^0}\right)\\
&\hspace{-2em}\displaystyle+b_k^{\mathrm{local}}\sum_{n=0}^{N+1}\sum_{\ell\in\mathcal{N}_k}\Big[\frac{\beta_{k\ell}^0-{u}_{k\ell}^\star+{\theta}_k(n)^\star\widehat{\beta}_{k\ell}}{\gamma_k}\Big]\\
 &\hspace{-2em}\displaystyle-c_k \sum_{n=0}^N\sum_{\ell\in\widehat{\mathcal{N}}_k(n)}\log(v_{k\ell}^{\star}(n))+d_k\sum_{n=0}^{N+1} {\theta}_k(n)^\star
\end{array} \] where $\forall k\in\mathcal{K}$, \[ [\boldsymbol{{U}^\star}]_{k\ell}={u}_{k\ell}^\star:=\left\{\begin{array}{l}
\beta_{k\ell}^0-\exp(\xi_{y_{k\ell}}^{\mathrm{GNE}})\text{ if }\ell\in\mathcal{N}_k\\
0\text{ otherwise,}
\end{array}\right.\]  $\forall n\in\{0,\ldots,N\}$,\\
$\displaystyle {\theta}_k(n+1)^\star=\sum_{\ell_n\in\widehat{\mathcal{N}}_k(n)}\sum_{\ell_{n-1}\in\widehat{\mathcal{N}}_{\ell_n}(n-1)}\ldots\sum_{\ell_0\in\widehat{\mathcal{N}}_{\ell_1}(0)}\Big[$
\[\hspace{3em}\alpha_{k\ell_n\ell_{{n-1}}\ldots,\ell_0}(n)\exp(\xi_{\omega_{k \ell_n \ell_{n-1} \ldots \ell_0}}^{\mathrm{GNE}}(n))\theta_{\ell_0}(0)\Big],\]
and $\forall \ell\in\widehat{\mathcal{N}}_k(n)$, $v_{k\ell}^{\star}(n):=$
\[\left\{\begin{array}{l}
\exp(\xi_{\omega_{k\ell k\ldots k}}^{\mathrm{GNE}}(n)-\xi_{\omega_{\ell k\ldots k}}^{\mathrm{GNE}}(n-1))\text{ if }n>0\\
\exp(\xi_{\omega_{k\ell}}^{\mathrm{GNE}}(0))\text{ otherwise.}
\end{array}\right.\]
Since $\xi^{\mathrm{GNE}}$ is a $\mathrm{GNE}$ of $\widetilde{\mathcal{G}}$, it follows from \eqref{eq:GNEgameaux} that, $\forall k\in\mathcal{K}$ and $\xi_k\in\boldsymbol{\Pi}_k\widetilde{C}$,
\[\widetilde{J}_k(\xi^{\mathrm{GNE}})\leq \widetilde{J}_k(\xi_k,\xi_{-k}^{\mathrm{GNE}})\] 
and  $\forall \xi_{K+1}\in\boldsymbol{\Pi}_{K+1}\widetilde{C}$,
\[\widetilde{J}_{K+1}(\xi^{\mathrm{GNE}})\leq \widetilde{J}_{K+1}(\xi_{K+1},\xi_{-(K+1)}^{\mathrm{GNE}}).\]
In view of \eqref{eq:rho}, it follows that:
\[
\begin{array}{ll}
J_k({u}^\star,{v}^\star)&=\widetilde{J}_k(\xi^{\mathrm{GNE}})+b_k^{\mathrm{global}}\widetilde{J}_{K+1}(\xi^{\mathrm{GNE}})\\
     &\leq \widetilde{J}_k(\xi_k,\xi_{-k}^{\mathrm{GNE}})+b_k^{\mathrm{global}}\widetilde{J}_{K+1}(\xi_{K+1},\xi_{-(K+1)}^{\mathrm{GNE}})\\
     & =J_k(u_k,v_k,{u}_{-k}^\star,{v}_{-k}^\star).
\end{array}
\]
Hence, $({u}^\star,{v}^\star)$ is a GNE of $\mathcal{G}$.\\

\noindent \emph{\textbf{\underline{Uniqueness of the $\mathrm{GNE}$:}}}

Let define a weighted non-negative sum of the function $\widetilde{J}_k$ given by $\sigma(\xi,\delta):=\sum_{k=1}^K \delta_k \widetilde{J}_k(\xi),\ \delta_k\in\R_{>0}$. Based on Rosen's theory of uniqueness \cite{rosen1965existence}, the following Definition is used for exhibiting the desired property of the equilibrium point.
 \begin{defn}
 $\sigma(\xi,\delta)$ is diagonally strictly convex ($\mathrm{DSC}$) for $\xi\in\mathcal{E}$ and fixed $\delta\in\Rlo^{K+1}$ if for every $\xi^0,\xi^1\in\widetilde{\mathcal{C}}$ we have \[(\xi^1-\xi^0)^\top(g(\xi^1,\delta)-g(\xi^0,\delta))>0,\] where $g(\xi,\delta):=[\delta_1 \nabla_{\xi_1} \widetilde{J}_1(\xi),\ldots,\delta_{K+1}\nabla_{\xi_{K+1}}\widetilde{J}_{K+1}(\xi)]^\top$.
 
 \hfill$\Box$
 \end{defn}
 In what follows, we make use of the following function: for $(\xi,\widehat{\xi})\in\widetilde{\mathcal{C}}^2$ and $\delta\in\Rlo^{K+1}$
\begin{equation}\label{eq:rhoGNE}
    \rho(\xi,\widehat{\xi},\delta):=\sum_{k=1}^{K+1} \delta_k\widetilde{J}_k(\xi_1,\ldots,\xi_{k-1},\widehat{\xi}_k,\xi_{k+1},\ldots,\xi_K).
\end{equation} 
 In the following we guarantee the uniqueness property of the $\mathrm{GNE}$ in the game $\widetilde{\mathcal{G}}$. In what follows, we denote by $M:=\mathrm{dim}(\widetilde{h}(\xi))$. The Kuhn-Tucker conditions that verify \eqref{eq:GNEgameaux} can now be expressed as follows: $ \forall k\in\mathcal{K},\ \exists \mu_k^{\mathrm{GNE}}\in\R_{\leq 0}^{M}$ such that,
\begin{subequations}
\begin{equation}\label{eq:primalcnd}
     \widetilde{h}(\xi^{\mathrm{GNE}})\leq 0
\end{equation}
\begin{equation}\label{eq:Complementaryslacknesscnd}
    (\mu_k^{\mathrm{GNE}})^\top \widetilde{h}(\xi^{\mathrm{GNE}})=0
\end{equation}
\begin{equation}\label{eq:Stationaritycnd}
    \delta_k\nabla_{\xi_k}\widetilde{J}_k(\xi^{\mathrm{GNE}})+ (\mu_k^{\mathrm{GNE}})^\top \nabla_{\xi}\widetilde{h}(\xi^{\mathrm{GNE}})=0.
\end{equation}
\end{subequations}
Let $\overline{\delta}\in\R_{>0}^{K+1}$. In view of the geometric properties of $\widetilde{J}_k$, it follows that $\rho(\xi,\widehat{\xi},\overline{\delta})$ is continuous in $\xi$ and $\widehat{\xi}$ and convex in $\widehat{\xi}$ for every fixed $\xi\in\mathcal{E}$.  From the Definition of the $\mathrm{DSC}$, we have for every $(\xi^0,\xi^1)\in\widetilde{\mathcal{C}}^2$,

\[\begin{array}{lll}
  &\hspace{-1em} (\xi^1-\xi^0)^\top(g(\xi^1,\overline{\delta})-g(\xi^0,\overline{\delta}))=\sum_{k=1}^K \overline{\delta}_k\Bigg[(N+2)\\
 &\hspace{-1em}\times\displaystyle\sum_{\ell=1}^K\Big[\frac{b_k^{\mathrm{local}}(\xi_{y_{k\ell}}^1-\xi_{y_{k\ell}}^0)\displaystyle(\exp{(\xi_{y_{k\ell}}^1)}-\exp{(\xi_{y_{k\ell}}^0)})}{\gamma_k}\Big]+\sum_{n=0}^{N}\Big[\\
 &\hspace{-1em}\displaystyle\Big[\sum_{\ell=1}^K\frac{b_k^{\mathrm{local}}\widehat{\beta}_{k\ell}\displaystyle}{\gamma_k}+d_k\Big]\times\sum_{\ell_n\in\widehat{\mathcal{N}}_k(n)}\hspace{-1em}\ldots\sum_{\ell_0\in\widehat{\mathcal{N}}_{\ell_1}(0)}\hspace{-1em}\big[\alpha_{k \ell_n \ldots \ell_0}(n)\times\\
 &\hspace{-1em}(\xi_{w_{k \ell_n \ldots \ell_0}}^{1}(n)-\xi_{w_{k \ell_n \ldots \ell_0}}^{0}(n))\Big(\exp{(\xi_{w_{k \ell_n  \ldots \ell_0}}^{1}(n))}-\\
 &\hspace{-1em}\displaystyle\exp{(\xi_{w_{k \ell_n \ldots \ell_0}}^{0}(n))}\Big)\theta_{\ell_0}(0)\big]\Big]\Bigg]+\overline{\delta}_{K+1}\times\\
 &\hspace{-1em}\displaystyle\sum_{n=0}^{N+1}\Big[(\xi_{\lambda }^{1}(n)-\xi_{\lambda }^{0}(n))(\exp{(\xi_{\lambda }^{1}(n))}-\exp{(\xi_{\lambda }^{0}(n))})\Big]>0\\
 &\hspace{-1em}\Rightarrow\sigma(\xi,\overline{\delta})\text{ is }\mathrm{DSC},\ \forall \xi\in\widetilde{\mathcal{C}}
 \end{array}\] Then by the Kakutani fixed point theorem, there exists $\xi^\star(\overline{\delta})\in\widetilde{\mathcal{C}}$ such that
\[\rho(\xi^\star(\overline{\delta}),\xi^\star(\overline{\delta}),\overline{\delta})=\min\limits_{\xi\in\widetilde{\mathcal{C}}}\rho(\xi^\star(\overline{\delta}),\xi,\overline{\delta}).\]
Then by the necessary $\widetilde{h}(\xi^\star(\overline{\delta}))\leq 0$, it follows that $\exists \mu^\star\in\R_{\leq 0}^{M}$ such that, ${\mu^\star}^\top \widetilde{h}(\xi^\star(\overline{\delta}))=0$ and $\forall k\in\mathcal{K}$,
\[\overline{\delta}_k\nabla_{\xi_k}\widetilde{J}_k(\xi^\star(\overline{\delta}))+\sum_{\ell=1}^{M}\mu_{\ell}^\star\nabla_{\xi_k}h_{\ell}(\xi^\star(\overline{\delta}))=0,\]
which are the conditions \eqref{eq:primalcnd}, \eqref{eq:Complementaryslacknesscnd} and \eqref{eq:Stationaritycnd} with $\xi^\star(\overline{\delta})=\xi^{\mathrm{GNE}}$ and $\forall k\in\mathcal{K}\cup\{K+1\},\ \ell\in\{1,\ldots,M\},\ \mu_{k\ell}^{\mathrm{GNE}}=\mu_{\ell}^\star/\overline{\delta}_k$, which are sufficient to ensure that $\xi^\star(\overline{\delta})$ is a GNE (i.e., $\xi^\star(\overline{\delta})$ verifies \eqref{eq:GNEgameaux}); according to \cite[Thm. 4]{rosen1965existence}, $\xi^\star(\overline{\delta})$ is a unique normalized equilibrium point for the specified value of $\delta=\overline{\delta}$.


\subsection{Proof of the Proposition \ref{Prop:Efficiency}}

\noindent Let $\xi^\star\in\argmin\limits_{\xi\in\widetilde{\mathcal{C}}}\sum_{k=1}^K[\widetilde{J}_k(\xi_1,\ldots,\xi_K)+b_k^{\mathrm{global}}\widetilde{J}_{K+1}(\xi_{K+1})]$ and $\xi\in\widetilde{\mathcal{C}}$. Let us denote by $\widetilde{\mathcal{C}}_{1:K}(\xi_{K+1}):=\{(\xi_1,\ldots,\xi_K):\widetilde{h}(\xi)\leq 0\}$ and $\widetilde{\mathcal{C}}_{K+1}(\xi_{1},\ldots,\xi_K):=\{\xi_{K+1}:\widetilde{h}(\xi)\leq 0\}$.  It follows that 
\[\begin{array}{ll}
      \hspace{-1em}\displaystyle\sum_{k=1}^K[\widetilde{J}_k(\xi_1,\ldots,\xi_K)+b_k^{\mathrm{global}}\widetilde{J}_{K+1}(\xi_{K+1})]\geq \hspace{-2.5em}\min\limits_{(\xi_1,\ldots,\xi_K)\in{\widetilde{\mathcal{C}}}_{1:K}(\xi_{K+1}^\star)}\\
     \displaystyle  \hspace{-0.5em}\big[ \sum_{k=1}^K\widetilde{J}_k(\xi_1,\ldots,\xi_K)\big]+b_k^{\mathrm{global}}\hspace{-1.5em}\min\limits_{\xi_{K+1}\in\widetilde{\mathcal{C}}_{K+1}(\xi_1^\star,\ldots,\xi_K^\star)}\hspace{-2.5em}\big[\widetilde{J}_{K+1}(\xi_{K+1})\big].
\end{array}\]
According to the Perron-Frobenius lemma and the change of variable in \eqref{eq:uV},  \[\begin{array}{l}
 \displaystyle\min\limits_{\xi_{K+1}\in\widetilde{\mathcal{C}}_{K+1}(\xi_1^\star,\ldots,\xi_K^\star)}\widetilde{J}_{K+1}(\xi_{K+1})\\
 \displaystyle=\sum_{n=0}^{N+1}\rho(\boldsymbol{\boldsymbol{D}_{\gamma}^{-1}}(\mathrm{Exp}(\xi_y^\star)+\mathrm{diag}(\widetilde{\theta}(n)^\star)\widehat{\boldsymbol{B}}),
\end{array}\] with  $\forall n\in\{0,\ldots,N\}$,\\
$\displaystyle \widetilde{\theta}_k(n+1)^\star=\sum_{\ell_n\in\widehat{\mathcal{N}}_k(n)}\sum_{\ell_{n-1}\in\widehat{\mathcal{N}}_{\ell_n}(n-1)}\ldots\sum_{\ell_0\in\widehat{\mathcal{N}}_{\ell_1}(0)}\Big[$
\[    \hspace{3em}\alpha_{k \ell_n \ell_{n-1} \ldots \ell_0}(n)\exp(\xi_{\omega_{k \ell_n \ell_{n-1} \ldots \ell_0}}(n)^\star)\theta_{\ell_0}(0)\Big].
\]Furthermore, \[\hspace{-0.5em}\min\limits_{(\xi_1,\ldots,\xi_K)\in{\widetilde{\mathcal{C}}}_{1:K}(\xi_{K+1}^\star)}  \sum_{k=1}^K\widetilde{J}_k(\xi_1,\ldots,\xi_K)=\sum_{k=1}^K\widetilde{J}_k(\xi_1^\star,\ldots,\xi_K^\star).\] Finally, we derive that $\displaystyle\sum_{k=1}^K\Bigg[\widetilde{J}_k(\xi_1^\star,\ldots,\xi_K^\star) +$
\[\begin{array}{ll}
     \displaystyle b_k^{\mathrm{global}}\sum_{n=0}^{N+1}\rho(\boldsymbol{\Gamma^{-1}}(\mathrm{Exp}(\xi_y^\star)+\mathrm{diag}(\widetilde{\theta}(n)^\star)\boldsymbol{B}^d)\Bigg] \\
     \displaystyle=\min\limits_{(u,v)\in\mathcal{C}}\sum_{k=1}^K J_{k}(u,v).
\end{array}
\]


\subsection{Proof of Proposition \ref{Prop:Determination}}
We recall that $M=\mathrm{dim}(\widetilde{h}(\xi))$ and in what follows we denote $\widetilde{M}:=\mathrm{dim}(\xi)$. Let $\overline{\delta}\in\R_{>0}^{K+1}$. Consider the following differential equations, $\forall k\in\mathcal{K}\cup\{K+1\}$,
\begin{equation}\label{eq:dyn_algok}
    \frac{d\xi_k}{dt}=-\overline{\delta}_k\nabla_{\xi_k}\widetilde{J}_k+\sum_{j=1}^{M}\mu_{j}\nabla_{\xi_k}\widetilde{h}_j(\xi),\ \mu\in\mathcal{M}(\xi),
\end{equation}where $\mathcal{M}(\xi)\subset\R_{\leq 0}^{M}$ is bounded. We define $\boldsymbol{H}:\widetilde{\mathcal{C}}\to\R^{\widetilde{M}\times M}$ by \[\boldsymbol{H}(\xi):=[\nabla_{\xi}\widetilde{h}_{1}(\xi),\nabla_{\xi}\widetilde{h}_{2}(\xi),\ldots,\nabla_{\xi}\widetilde{h}_{M}(\xi)].\] The matrix formulation of \eqref{eq:dyn_algok} is given by:
\begin{equation}\label{eq:dyn_algo}
\frac{d \xi}{dt}=f(\xi,\mu,\overline{\delta}):=-g(\xi,\overline{\delta})+\boldsymbol{H}(\xi)\mu,\ \mu\in\mathcal{M}(\xi),
\end{equation}with 
\[\begin{array}{ll}
     &  \mathcal{M}(\xi):=\hspace{0.5em}\argmin\limits_{\mu}\Vert f(\xi,\mu,\overline{\delta})\Vert\\
     &  \begin{array}{lll}
          &  \text{s.t.}& \begin{cases}
               \mu_j\leq 0 \text{ if }\widetilde{h}_j(\xi)>0\\
               \mu_j=0\text{ otherwise.}
          \end{cases}
     \end{array}
     \end{array}\]
According to \cite[Thm. 7]{rosen1965existence}, for every starting point $\xi(0)\in\widetilde{\mathcal{C}}$, the trajectory $\xi(t)$ of \eqref{eq:dyn_algo} exists and remains in $\widetilde{\mathcal{C}}$ at any time $t>0$. In what follows, we show the convergence of \eqref{eq:dyn_algo} to the unique normalized equilibrium point $\xi^\star(\overline{\delta})$ associated to the value $\overline{\delta}$. 

We consider an equilibrium point ${\xi}^\star$ of system \eqref{eq:dyn_algo} for a fixed $\overline{\delta}\in\R_{>0}^{K+1}$ such that $f({\xi}^\star,\mu,\overline{\delta})=0,\ \mu\in\mathcal{M}({\xi}^\star)$. From the proof of Proposition \ref{Prop:Existence_Uniqueness} and the definition of $f$, for $\xi^\star$ and $\mu\in\mathcal{M}(\xi^\star)$ such that $f(\xi^\star,\mu,\overline{\delta})=0$ is obviously a normalized equilibrium point associated to the fixed value $\overline{\delta}$. For all $\xi\in\widetilde{\mathcal{C}}$, we define $\overline{\mu}(\xi)$ such as the nonzeros elements of of $\mu\in\mathcal{M}(\xi)$ which are given by $\overline{\mu}\in \R_{\leq 0}^{\overline{M}}$, where $\overline{M}\leq M $ and $\overline{\mu}(\xi)=\left[\boldsymbol{\overline{H}}(\xi)^\top \boldsymbol{\overline{H}}(\xi)\right]^{-1}\boldsymbol{\overline{H}}(\xi)^\top g(\xi,\overline{\delta})\leq 0$, where the matrix $\boldsymbol{\overline{H}}(\xi)\in\R^{\widetilde{M}\times \overline{M}}$ is composed by $\overline{M}\leq M $ linearly independent columns of $\boldsymbol{H}(\xi)$ selected from $\nabla_{\xi} \widetilde{h}_j(\xi)$ for $j\in\{i\in\{1,\ldots,M\} :\ \widetilde{h}_i(\xi)>0\}$. It follows that: 
\begin{equation}\label{eq:dfdt}\frac{d f(\xi,\mu,\overline{\delta})}{dt}=\left(-\boldsymbol{G}(\xi,\overline{\delta})+\sum_{j=1}^{M}\mu_{j}\boldsymbol{Q}_{j}(\xi)\right)\frac{d \xi}{dt}+\boldsymbol{\overline{H}}(\xi)\frac{d \overline{\mu}}{dt},
\end{equation}
where $\boldsymbol{Q}_j(\xi)$ is the hessian of $\widetilde{h}_j(\xi)$; $\boldsymbol{G}(\xi,\overline{\delta})$ is the jacobian of $g(\xi,\overline{\delta})$; which are both positive semi-definite since $\forall k,\ \widetilde{J}_k$ and $h$ are convex w.r.t. $\xi$. Let us consider $V:\widetilde{\mathcal{C}}\times\Rlo^{M}\to\Rlo$ as a Lyapunov function, given by $V(\xi,\mu)=\frac{1}{2}\Vert f(\xi,\mu,\overline{\delta})\Vert^2$, which is continuously differentiable positive definite function on $\widetilde{\mathcal{C}}\times\R_{\leq 0}^{M}$. By combining \eqref{eq:dyn_algo} with \eqref{eq:dfdt}, we derive that,
\[\begin{array}{ll}
     \frac{d}{dt}V(\xi,\mu)&=\frac{1}{2}\frac{d }{dt}(f(\xi,\mu,\overline{\delta})^\top f(\xi,\mu,\overline{\delta}))\\
     & =f(\xi,\mu,\overline{\delta})^\top\frac{d}{dt}f(\xi,\mu,\overline{\delta})\\
     &=-f(\xi,\mu,\overline{\delta})^\top\boldsymbol{G}(\xi,\overline{\delta})f(\xi,\mu,\overline{\delta})\\
     &+\hspace{-3.5em}\sum\limits_{j\in\{1\leq i\leq M:\ \widetilde{h}_i(\xi)>0\}}\hspace{-3.5em} \overline{\mu}_j f(\xi,\mu,\overline{\delta})^\top \boldsymbol{Q}_j(\xi)f(\xi,\mu,\overline{\delta})\\
     &+f(\xi,\mu,\overline{\delta})^\top \boldsymbol{\overline{H}}(\xi)\frac{d\overline{\mu}}{dt}.
\end{array}\] From \eqref{eq:dyn_algo} and the expression of $\overline{\mu}$, it follows that,

$\displaystyle f(\xi,\mu,\overline{\delta})^\top\boldsymbol{\overline{H}}(\xi)\frac{d\overline{\mu}}{dt}$
\[\begin{array}{ll}
      =&\displaystyle\left[-g(\xi,\overline{\delta})^\top\boldsymbol{\overline{H}}(\xi)+\overline{\mu}^\top\boldsymbol{\overline{H}}(\xi)^\top\boldsymbol{\overline{H}}(\xi)\right]\frac{d\overline{\mu}}{dt}\\
      =&\displaystyle\left[-g(\xi,\overline{\delta})^\top\boldsymbol{\overline{H}}(\xi)+g(v,\overline{\delta})^\top\boldsymbol{\overline{H}}(\xi)\right]\frac{d\overline{\mu}}{dt}\\
      =&0.
\end{array}
\]
Since $\boldsymbol{Q_j}(\xi)$ and $\boldsymbol{G}(\xi,\overline{\delta})$ are positive semidefinite, it follows that, $\displaystyle\frac{d}{dt}V(\xi,\mu)$
\[\begin{array}{ll}
&=f(\xi,\mu,\overline{\delta})^\top\left[-\boldsymbol{G}(\xi,\overline{\delta})\right]f(\xi,\mu,\overline{\delta})\\
&+\hspace{-3.5em}\sum\limits_{j\in\{1\leq i\leq M:\ \widetilde{h}_i(\xi)>0\}}\hspace{-3.5em} \overline{\mu}_j f(\xi,\mu,\overline{\delta})^\top \boldsymbol{Q_j}(\xi)f(\xi,\mu,\overline{\delta})\leq 0.
\end{array}\] Let $\mathcal{S}:=\{(\xi,\mu)\in\widetilde{\mathcal{C}}\times\Rlo^{M}:\ \frac{d}{dt}V(\xi,\mu)=0\}$. Since, for a fixed $\overline{\delta}$, there exists a unique normalized equilibrium point $\xi^\star(\overline{\delta})$ that verifies the optimization problem given in 
 \[\xi^\star(\overline{\delta})=\min\limits_{\xi\in\widetilde{\mathcal{C}}}\rho(\xi^\star(\overline{\delta}),\xi,\overline{\delta}),\] where $\rho$ is given in \eqref{eq:rhoGNE}. It follows that no solution of system \eqref{eq:dyn_algo} can stay identically in $\mathcal{S}$ other than the solution $\xi(t)=\xi^\star(\overline{\delta})$. Then, according to \cite[Corollary 4.1]{Khalil_02}, for any initial condition $\xi(t=0)\in\widetilde{\mathcal{C}}$, the system \eqref{eq:dyn_algo} converge asymptotically to the normalized equilibrium point $\xi^\star(\overline{\delta})$, which is the unique GNE of $\widetilde{\mathcal{G}}$.


\bibliographystyle{unsrt}
\bibliography{IEEEabrv,main}

\end{document}

%% file: commands.tex
\newcommand{\R}{\ensuremath{\mathbb{R}}}

\newcommand{\Rlo}{\ensuremath{\mathbb{R}_{\geq 0}}}

\definecolor{bleucit}{rgb}{0.2,0.4,0.6} 

\newcommand\BLUE[1]{{\color{black}#1}}

\definecolor{blue_cv}{rgb}{0.09,0.35,0.78}

\newcommand{\Argmin}{\ensuremath{\text{argmin}\,}}
\DeclareMathOperator*{\argmin}{\Argmin}

\newtheorem{defn}{Definition}

\newtheorem{ass}{Assumption}

\newtheorem{prop}{Proposition}